\documentclass{dukecei}

\usepackage[utf8]{inputenc}
\usepackage{microtype}
\usepackage{longtable}
\usepackage{tikz,pgfplots}
\pgfplotsset{compat=1.18}
\usetikzlibrary{arrows.meta,positioning,fit,patterns,decorations.pathreplacing}
\usepgfplotslibrary{groupplots}
\definecolor{methodMaroon}{HTML}{800000}
\definecolor{baselineSky}{HTML}{87CEEB}
\definecolor{baselineLight}{HTML}{ADD8E6}
\definecolor{baselineCyan}{HTML}{E0FFFF}
\definecolor{baselineLine}{HTML}{36738A}
\definecolor{figureInk}{HTML}{252525}
\definecolor{figureGrid}{HTML}{DEDEDE}
\newcommand{\plotfont}{\sffamily\fontsize{8}{9.2}\selectfont}
\pgfplotsset{
 paperaxis/.style={
  font=\plotfont,label style={font=\plotfont},tick label style={font=\plotfont},
  title style={font=\plotfont},axis line style={black!65,line width=.4pt},
  tick style={black!65,line width=.4pt},tick align=outside,
  grid=major,grid style={figureGrid,densely dotted,line width=.25pt},
  legend style={font=\plotfont,draw=none,fill=white,cells={anchor=west},inner sep=2pt},
  line width=.85pt,mark size=2.2pt,
  every axis plot/.append style={line width=.85pt},
  scaled ticks=false,
 },
 uniformline/.style={draw=baselineLine,mark=square*,mark options={fill=baselineSky,draw=baselineLine}},
 splitmarks/.style={only marks,mark=diamond*,draw=methodMaroon,fill=methodMaroon,mark size=3pt},
 fullbar/.style={draw=baselineLine,fill=baselineSky,postaction={pattern=north east lines,pattern color=baselineLine!45}},
 splitbar/.style={draw=methodMaroon,fill=methodMaroon},
}
\tikzset{
 paperbox/.style={draw=figureInk,line width=.5pt,rounded corners=1.2pt,align=center,font=\plotfont,inner sep=4pt},
 paperarrow/.style={-{Stealth[length=1.5mm]},draw=figureInk,line width=.65pt},
}

\usepackage{xurl}

\newlength{\ceiplotwidth}
\title{Not All Loops Cost the Same: Phase-Dependent Depth for Serving Looped Transformers}

\author{
    Zhixu Du$^{1}$\footnote{Correspondence E-mail: zhixu.du@duke.edu},
    Yuhan Cheng$^{1}$,
    Hai (Helen) Li$^{1}$,
    Yiran Chen$^{1}$\\[1em]
    \normalsize $^{1}$Department of Electrical and Computer Engineering, Duke University, Durham, NC, USA
}

\begin{document}

\maketitle
\thispagestyle{firstpagestyle} 

\begin{abstract}
Looped transformers scale inference-time reasoning by repeatedly reusing a shared core, increasing computation without increasing parameter count. Serving prefilling and decoding with a uniform loop budget couples their costs. We present \textit{LoopDuet}, a phase-dependent depth and training-free serving approach that selects the two budgets jointly while preserving decoding depth at prefilling boundaries. Our vLLM implementation gathers tokens requiring further recurrence into compact batches, turning the selected allocation into executable savings with existing weights and attention kernels. At runtime, a queue-age controller adjusts prefilling depth to retain quality as demand changes. Experiments on the full HotpotQA dataset with one A100-80GB demonstrate two benefits of this flexibility. First, separate phase budgets improve the quality--throughput trade-off: Huginn delivers 19\% more throughput than the fastest tested uniform setting meeting the same accuracy criterion, while a higher-throughput operating point gains 6.04 EM points over uniform depth eight at 97\% of its throughput. Second, the same execution mechanism retains more quality under load: on two evaluated OLMo arrival traces, adaptive prefilling gains 0.92--1.35 EM points over the most accurate tested fixed-depth controls at the same 100\% attainment of a one-second time-to-first-token target. Together, these results establish LoopDuet as a practical way to turn flexible model computation into higher serving capacity and better quality under load.
\end{abstract}

\section{Introduction}
Looped transformers scale reasoning computation by repeatedly applying a shared core, increasing
effective depth without increasing parameter count \citep{geiping2025scaling,mcleish2026teaching}. Shared weights support
different inference budgets, offering quality--cost choices at deployment. 
Exploiting this flexibility requires a \textit{depth allocation}: a rule that decides how many loops each token receives.

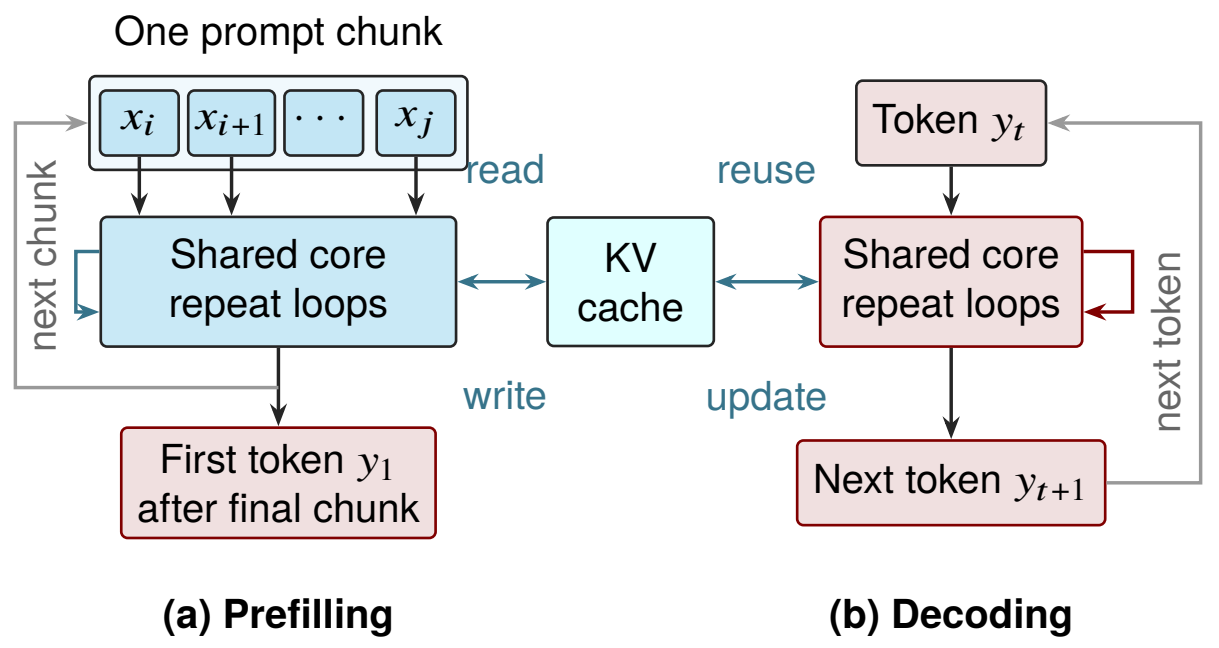
\begin{figure}[t]
\centering
\vspace{1em}
\begin{tikzpicture}[x=1cm,y=1cm,font=\plotfont]
\path[use as bounding box] (0,0) rectangle (8.05,4.05);
\tikzset{overviewcore/.style={paperbox,minimum width=1.65cm,minimum height=.85cm},
 overviewtoken/.style={paperbox,minimum height=.56cm,inner sep=3pt}}
\node[anchor=south] at (1.80,3.72) {One prompt chunk};
\node[paperbox,fill=baselineSky!12,minimum width=2.50cm,minimum height=.62cm,inner sep=0pt] (chunk) at (1.80,3.40) {};
\foreach \x/\lab/\id in {.88/{x_i}/a,1.49/{x_{i+1}}/b,2.10/{\cdots}/c,2.71/{x_j}/d} {
 \node[paperbox,fill=baselineSky!50,minimum width=.52cm,minimum height=.43cm,inner sep=1pt] (prompt-\id) at (\x,3.40) {$\lab$};
}
\node[overviewcore,fill=baselineSky!45,minimum width=2.35cm] (prefill) at (1.80,2.35) {Shared core\\repeat loops};
\foreach \id in {a,b,d} {
 \draw[paperarrow] (prompt-\id.south)--(prompt-\id.south |- prefill.north);
}
\node[overviewtoken,draw=methodMaroon,fill=methodMaroon!12,align=center] (first) at (1.80,1.02) {First token $y_1$\\after final chunk};
\draw[paperarrow] (prefill.south)--(first.north);
\draw[paperarrow,baselineLine] ([yshift=2mm]prefill.west)--(.46,2.55)--(.46,2.15)--([yshift=-2mm]prefill.west);
\draw[paperarrow,gray!80] (1.80,1.65)--(.06,1.65)--(.06,3.40)--(chunk.west);
\node[rotate=90,text=gray!80!black,inner sep=1pt] at (.23,2.52) {next chunk};
\node[overviewtoken,fill=methodMaroon!12] (token) at (6.25,3.40) {Token $y_t$};
\node[overviewcore,draw=methodMaroon,fill=methodMaroon!12] (decode) at (6.25,2.35) {Shared core\\repeat loops};
\node[overviewtoken,draw=methodMaroon,fill=methodMaroon!12] (next) at (6.25,1.02) {Next token $y_{t+1}$};
\draw[paperarrow] (token.south)--(decode.north);
\draw[paperarrow] (decode.south)--(next.north);
\draw[paperarrow,methodMaroon] ([yshift=2mm]decode.east)--(7.45,2.55)--(7.45,2.15)--([yshift=-2mm]decode.east);
\draw[paperarrow,gray!80] (next.east)--(7.90,1.02)--(7.90,3.40)--(token.east);
\node[rotate=90,anchor=south,text=gray!80!black] at (7.90,1.95) {next token};
\node[paperbox,fill=baselineCyan,minimum width=1.10cm,minimum height=.85cm] (kv) at (4.13,2.35) {KV\\cache};
\draw[{Stealth[length=1.4mm]}-{Stealth[length=1.4mm]},draw=baselineLine,line width=.65pt] (prefill.east)--(kv.west);
\node[anchor=south,text=baselineLine] at (3.30,2.87) {read};
\node[anchor=north,text=baselineLine] at (3.30,1.83) {write};
\draw[{Stealth[length=1.4mm]}-{Stealth[length=1.4mm]},draw=baselineLine,line width=.65pt] (kv.east)--(decode.west);
\node[anchor=south,text=baselineLine] at (5.03,2.87) {reuse};
\node[anchor=north,text=baselineLine] at (5.03,1.83) {update};
\node[anchor=north,font=\plotfont\bfseries] at (1.80,.39) {(a) Prefilling};
\node[anchor=north,font=\plotfont\bfseries] at (6.25,.39) {(b) Decoding};
\end{tikzpicture}
\caption{\textbf{Prefilling advances by chunk; decoding advances by token.} Prefilling processes chunk tokens in parallel. Cached context carries across chunks; the final chunk produces the first output token. Both phases reuse the same core weights.}
\label{fig:loop_overview}
\end{figure}

LLM serving processes each request in two phases, prefilling and decoding (Figure~\ref{fig:loop_overview}), which provide a natural basis for depth allocation. 
Prefilling processes prompt tokens together in several chunks, to build the KV cache; the final prompt position produces the first output token.
Decoding then generates subsequent tokens one position at a time. A uniform loop budget couples
their costs: lowering prefilling depth also lowers decoding depth. For long prompts and short
answers, retaining deep decoding therefore incurs repeated computation across the entire prompt.

Prior work exploits a common asymmetry between the phases: decoding is more sensitive to reduced computation than prefilling \citep{he2026popprefillonlypruningefficient}. 
SwiftKV \citep{qiao-etal-2025-swiftkv} computes prompts' later-layer KV cache directly from an earlier layer's output, skipping some later-layer computation during prefilling while decoding runs the full model; because this changes the input to the later layers' KV projections, it distills their Q,K, and V projections from the original model. 
POP \citep{he2026popprefillonlypruningefficient} applies a similar early exit during prefilling without training, but skips fewer layers since more aggressive pruning degrades quality. 
SPEED \citep{oh2026shallowprefilldeepdecoding} processes prefilling tokens through only the first K layers and omits their deep-layer KV states, reducing both prefilling computation and decode-time KV reads, but requires fine-tuning. 
Our work applies this asymmetry to looped transformers without training: it reduces prefilling loops rather than physical layers, selects prefilling and decoding depths jointly, and adapts prefilling depth to load during serving.

We present LoopDuet, a training-free serving approach that jointly selects prefilling depth $P$ and
decoding depth $S$, with $P\leq S$. Joint selection includes both uniform budgets and split budgets
that retain deeper decoding. In a split setting, ordinary prefilling positions stop after $P$ loops,
while decoding positions and scheduled prompt-chunk endpoints receive $S$ (Section \ref{subsec:separate_prefilling_decoding_depth}). This preserves the final
prompt position's first-token computation while reducing recurrent work over the surrounding
context.

This allocation saves computation only if the engine stops precessing positions that have finished their P loops. In a serving batch, however, each engine step mixes prompt chunks and decoding tokens from many requests, and after P loops only one query per request still needs recurrence. To remove completed prefilling positions from the subsequent core calls, LoopDuet applies a common loop boundary: after P loops, our vLLM implementation gathers the positions that need further recurrence into a compact batch for the remaining $S-P$ loops. Prompt context remains available in a shared KV cache, so the compact queries can attend to it using existing attention kernels and weights (Section \ref{sec:implementation}).

A fixed P either misses first-token latency targets during bursts or gives up quality when capacity is available. To track changing request rates, LoopDuet starts from the jointly selected operating point and uses a queue-age controller that lowers $P$ when first-token backlog grows and restores it when capacity permits (Section \ref{sec:controller_method}).

On the full
HotpotQA dataset with one A100-SXM4-80GB, Huginn $(S,P)=(32,16)$ delivers 19\% more
throughput than the fastest tested uniform setting meeting the same accuracy criterion. At a
higher-throughput point, $(14,8)$ gains 6.04 exact-match (EM) points over uniform depth eight while
retaining 97\% of its throughput.
This flexibility also retains more quality under load. For OLMo on HotpotQA, joint selection identifies
uniform depth four as the fastest tested setting meeting the accuracy criterion. Starting there, the
controller keeps $S=4$ and adjusts $P\in\{2,3,4\}$. On the evaluated 60 requests/s and 10/60
requests/s traces, it retains 0.92 and 1.35 more EM points than the most accurate tested fixed
prefilling depth with the same 100\% attainment of a one-second time-to-first-token (TTFT) target.

We make three contributions:
\begin{itemize}
\item \textbf{Joint depth selection:} We characterize quality--throughput choices by comparing phase-dependent allocations with competitive reduced uniform depths.
\item \textbf{Batched execution and runtime control:} We implement compact recurrence and queue-age adaptation in vLLM, translating depth allocation into serving capacity and quality retention under load.
\item \textbf{Role-dependent sensitivity:} We characterize depth sensitivity across two model families and three tasks, using recurrent-state measurements to screen candidate prefilling depths for task-quality validation.
\end{itemize}

\section{Serving Looped Transformers}
\label{sec:background}
\paragraph{Depth and memory are separate resources.}
The models we study consist of a prelude, a recurrent core, and a coda. The prelude computes an
input representation $e_i$ for token $i$. Each core application updates its recurrent state using
the previous state, $e_i$, and causal attention context; the coda produces the representation for
next-token prediction. Huginn uses 2/4/2 prelude/core/coda layers, while the OLMo retrofits use
4/6/4. Weight reuse controls parameter memory, but each loop incurs attention and feed-forward
computation.

Recurrent depth can also affect KV storage. A depth-indexed layout stores context for each
iteration, whereas terminal or shared layouts reuse a cache across loops, making memory scale with
physical layers rather than unrolled depth \citep{geiping2025scaling,huang2026loopedmodelsrightii,mishra2026enablingdynamiccomputationlooped}. Our serving
implementation uses one shared KV plane per core layer. Reducing prefilling depth removes
computation beyond this cache-sharing benefit. All serving comparisons use the same shared layout.

\paragraph{Prefilling work can dominate the arithmetic.}
Let $a$, $b$, and $c$ denote the numbers of prelude, core, and coda layers, and let $c_k = a + bk + c$ be the layer applications for a position receiving $k$ loops. For a prompt of length $L$ followed by $D$ decoding-token evaluations, uniform execution at depth $S$ performs $(L+D)\,c_S$ layer applications. 
For OLMo HotpotQA, prompts average 796 tokens and answers about 4.7 tokens, so prompt positions account for over 99\% of this work. This concentration motivates reducing prefilling work while retaining decoding depth.
Arithmetic savings and wall-time savings differ because prefilling processes many positions in
parallel, whereas decoding advances sequentially. Chunked prefilling and continuous batching further
couple their latency \citep{298679,kwon2023efficient}. The layer model identifies removable computation; phase
timing determines its wall-time value.

\begin{figure}[t]
\centering
\includegraphics[width=1.0\linewidth]{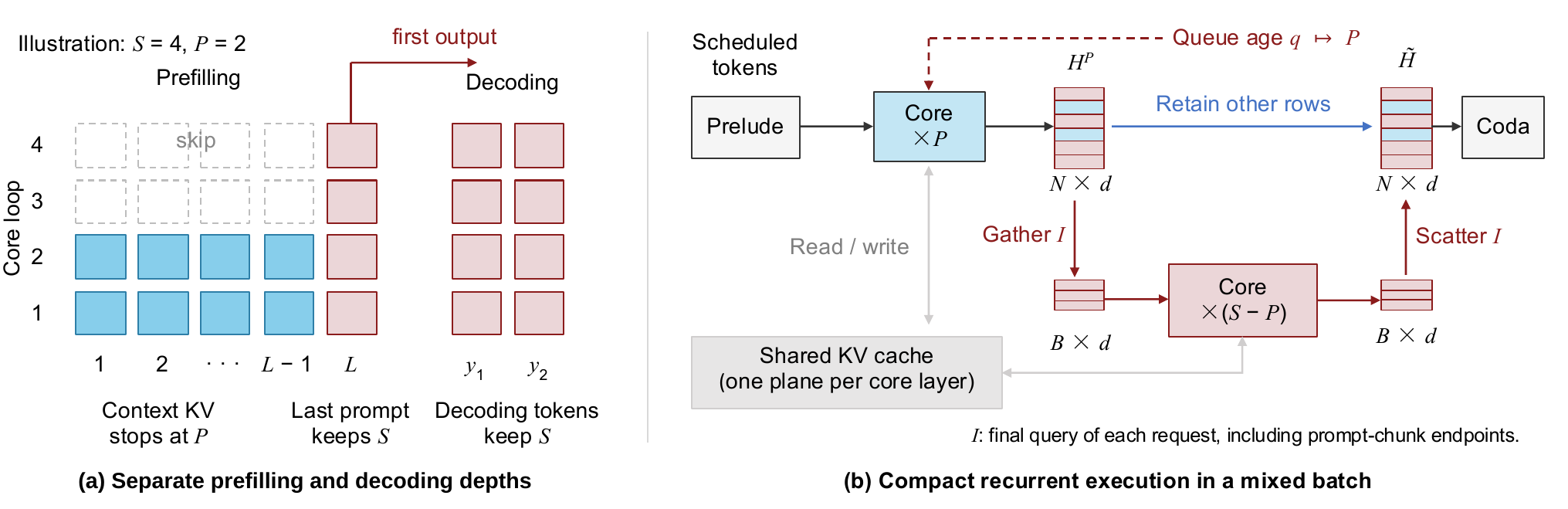}

\caption{\textbf{LoopDuet reduces prefilling loops while preserving decoding depth.}
(a) Ordinary prompt tokens receive $P$ loops during prefilling; the final prompt token and decoding tokens retain $S$.
(b) After $P$ loops on $N$ scheduled tokens, gather the final query of each of $B$ requests, run $S-P$ loops on this compact tensor, and scatter the updated rows before the coda. Here $d$ is the hidden width. Both paths use the same KV planes and existing attention kernels. The diagram shows hidden-state flow; prelude states and attention metadata use the same gather indices. The dashed arrow changes $P$ between engine steps. Every prompt-chunk endpoint retains $S$, including intermediate endpoints; single-query batches follow the full-depth path.}
\label{fig:framework}
\end{figure}

\paragraph{The optimization unit should survive batching.}
Conditional computation saves hardware work when per-token selections remove operations from the
executed batch. MoE systems illustrate this distinction between per-token selection and aggregate
expert traffic \citep{shazeer2017,gupta2026lynxenablingefficientmoe}. We use a common prefilling-loop boundary, after which the engine
gathers endpoints and decoding positions for further recurrence. 

\section{LoopDuet Design}
\label{sec:method}

LoopDuet allocates recurrent computation by serving phase. We first select prefilling and decoding
budgets jointly from their measured quality--throughput trade-off. The execution path then enforces
these budgets by gathering positions that require further recurrence, while retaining context from
positions that stop earlier. Finally, queue-age control adjusts prefilling depth within an
offline-characterized range. Figure~\ref{fig:framework} connects the allocation rule to this
execution path.

\subsection{Separate Prefilling and Decoding Depths}
\label{subsec:separate_prefilling_decoding_depth}
We write $(S,P)$ for decoding and prefilling core iterations, respectively, with $P\leq S$. For an
unchunked prompt of length $L$, positions $1,\ldots,L-1$ execute $P$ iterations. Position $L$
remains part of prefilling but executes $S$ iterations to produce the first output token. Every
subsequent generated position also executes $S$. Prelude and coda computation remain present for all
positions.

Continuing positions need access to the shallower prompt context. In the accuracy harness's
depth-indexed cache, skipped entries are filled from the last computed recurrence at the same
physical core layer. In the serving implementation's shared layout, each recurrence overwrites that
layer's KV plane, leaving each prompt position's last computed state available to later queries.
Decoding can therefore access prompt context without executing the omitted prefilling loops. We
evaluate task accuracy under this approximation to full-depth context.


\paragraph{Prefilling endpoints retain decoding depth.}
The final prompt position belongs to prefilling, but its coda output produces
the first output token. Reducing its depth would compute that token from a
shallower state; for short-answer tasks, the first token often determines the
answer. LoopDuet therefore keeps this position at $S$ while reducing work over
the preceding positions; Section~\ref{sec:accuracy} quantifies this choice. POP
likewise preserves first-token computation in its prefill-specific allocation
\citep{he2026popprefillonlypruningefficient}.

\subsection{Compact Execution with Shared Context}
\label{sec:implementation}

To realize these savings, the engine removes completed prefilling positions from later core calls.
For a batch of $N$ tokens from $B$ requests with hidden width $d$, the prelude and first $P$ core applications operate on the full batch. Tokens within each prompt chunk are processed with causal attention; successive loops update their recurrent states.

Let $I$ contain each request's final scheduled query index. Gathering these rows produces a $B\times
d$ tensor for the remaining $S-P$ loops. The corresponding prelude states, positions, and
one-query-per-request attention metadata follow the same indices. The gather reduces the active
queries while leaving prompt KV entries in the shared cache; continuing positions can therefore
attend to the full retained context through existing attention kernels. After recurrence, the engine
scatters the updated rows back into the $N\times d$ tensor before the coda. Other rows retain their
loop-$P$ states.

With chunked prefilling, \emph{every scheduled prompt-chunk endpoint} continues to depth $S$,
including intermediate endpoints. This gives each engine step the same final-query metadata rule;
only the final prompt endpoint produces the first output token. A batch whose maximum query length
is one follows the full-depth path. The supplement provides engine-step pseudocode.

Batches containing prefilling tokens run eagerly because the gathered tensor has a variable shape;
decoding-only batches use full CUDA graphs. Uniform-depth controls use the same execution modes. At
$P=S$, this configuration is already $1.42\times$ faster for OLMo and $1.22\times$ for Huginn than
vLLM's compiled mode in the recorded controls, with EM differences within 0.06 points. Our depth
comparisons therefore use the faster measured configuration.

\subsection{Adapt Prefilling Depth to Queue Age}
\label{sec:controller_method}
Static depth selection identifies a budget for a workload, but capacity requirements change with arrival rates. LoopDuet responds by adjusting $P$ while retaining the selected decoding depth. Let $q$ be the
age of the oldest admitted request awaiting its first token, including time in service. Before each
engine step, the controller computes
\begin{align}
 z &= \operatorname{clip}\!\left(\frac{q-0.25}{1.00-0.25},0,1\right),\\
 P &= \operatorname{round}\big(P_{\max}-z(P_{\max}-P_{\min})\big).
 \label{eq:control}
\end{align}
For an empty waiting set, $q=0$. Rising queue age lowers $P$ toward $P_{\min}$; receding backlog restores it toward $P_{\max}$. Decoding depth remains fixed.

For the main OLMo experiment, the accuracy–throughput comparison favors uniform depth four. We keep $S=4$ and
use $(P_{\max},P_{\min})=(4,2)$, with parameters fixed before all evaluated traces. Here $P_{\max}$
is the accuracy-screened starting depth, while $P_{\min}$ sets the quality--capacity trade-off
available under pressure. A prompt spanning multiple engine steps may encounter several values of
$P$.

Offline quality measurements establish this operating range; the runtime has no correctness
feedback. We therefore evaluate both task quality and TTFT attainment as the controller changes
depth. The appendix reports the supplementary $S=11$ control study separately.

\section{Evaluation}
\label{sec:evaluation}
The evaluation of LoopDuet follows the serving decision from joint depth selection (Section~\ref{sec:frontier})
to compact execution (Section~\ref{sec:serving}) and quality retention under load
(Section~\ref{sec:adaptive}). Role ablations and recurrent-state measurements then explain depth
sensitivity and assess candidate-budget screening (Sections~\ref{sec:accuracy}--\ref{sec:distance}).

\subsection{Setup and Comparison Protocol}
\label{sec:setup}
\paragraph{Models and tasks.}
We evaluate Huginn-0125 and recurrent OLMo-2 checkpoints trained at 16 and 32 loops
\citep{geiping2025scaling,mcleish2026teaching,walsh2025}. Their widths are 5,280 and 2,048, respectively. The accuracy study
uses the complete GSM8K test set, HotpotQA distractor dataset, and
ARC-Challenge (ARC) test set \citep{cobbe2021trainingverifierssolvemath,yang-etal-2018-hotpotqa,clark2018thinksolvedquestionanswering}. GSM8K uses five demonstrations and greedy
generation, with the last numeric answer extracted by the harness. ARC uses five demonstrations and
scores the choice labels at the first answer position.
HotpotQA requests short-phrase answers. OLMo uses one demonstration and a four-paragraph
construction retaining the two supporting paragraphs and two distractors, with distractors removed
as needed to meet a 1,000-token prompt cap. Huginn uses two demonstrations and a ten-paragraph
construction. These are within-model comparisons under fixed prompt construction, rather than an
accuracy ranking between models. The appendix records context variants and extraction details.

\paragraph{Metrics.}
We report task accuracy, with exact match (EM) and token-level F1 for
HotpotQA. All score differences are expressed in percentage points.
Joint-depth comparisons use $(32,32)$ as the baseline; prefilling-depth
ablations compare configurations at the same decoding depth $S$.

\paragraph{Serving.}
All timings use vLLM 0.14.1, bf16, one exclusively used A100-SXM4-80GB and no prefix caching. Wall time excludes model loading and an eight-request
warm-up. Closed-loop runs keep up to $C$ requests in flight; open-loop runs admit seeded Poisson or
alternating-rate arrivals when due. TTFT starts at scheduled arrival and includes queueing. Time per
output token (TPOT) is the interval from first to last token divided by the number of subsequent
tokens, evaluated on multi-token responses. Latency distributions describe variation within recorded
runs.

\paragraph{Evaluation implementation.}
Serving accuracy and performance are measured in the same vLLM runs.
We use a separate accuracy harness for token-role and KV-cache ablations.
For recurrent-state analysis, we first generate answers at each
checkpoint's reference depth. We then run the model on the original
prompts and recorded answers, keeping the answer tokens fixed and
recording hidden states at each loop.

\subsection{Joint Depth Selection Improves the Trade-off}
\label{sec:frontier}
\begin{figure}[t]\centering
\begin{tikzpicture}\begin{axis}[paperaxis,width=.99\ceiplotwidth,height=5.0cm,xmin=.7,xmax=3.95,ymin=22.7,ymax=35.8,xlabel={Throughput (req/s) $\uparrow$},ylabel={Exact match (\%) $\uparrow$},label style={font=\plotfont},tick label style={font=\plotfont},legend style={font=\plotfont,draw=none,at={(.02,.06)},anchor=south west},axis lines=box]
\addplot[uniformline] table[x=req_s,y=em,col sep=comma]{figures/frontier_huginn_uniform.csv};\addlegendentry{Uniform $(S,S)$}
\addplot[splitmarks] table[x=req_s,y=em,col sep=comma]{figures/frontier_huginn_split.csv};\addlegendentry{LoopDuet $(S,P)$}
\draw[densely dashed,gray] (axis cs:1.698,32.72)--(axis cs:1.822,33.69);
\draw[densely dashed,gray] (axis cs:3.587,24.29)--(axis cs:3.489,30.33);
\node[font=\plotfont,anchor=south west] at (axis cs:.976,33.79) {$(32,32)$};
\node[font=\plotfont,anchor=south west,text=methodMaroon] at (axis cs:1.822,33.69) {$(32,16)$};
\node[font=\plotfont,anchor=north east] at (axis cs:1.536,33.15) {$(20,20)$};
\node[font=\plotfont,anchor=south east,text=methodMaroon] at (axis cs:3.489,30.33) {$(14,8)$};
\node[font=\plotfont,anchor=north east] at (axis cs:3.587,24.29) {$(8,8)$};
\end{axis}\end{tikzpicture}
\caption{\textbf{Separate budgets improve Huginn's quality--throughput choices.} Full-set HotpotQA, one A100-80GB, $C=64$. LoopDuet $(32,16)$ gives $1.19\times$ throughput relative to the fastest tested uniform setting meeting the accuracy criterion, $(20,20)$. LoopDuet $(14,8)$ gains 6.04 EM points over uniform eight at 97\% of its throughput. Lines join measured uniform points as a visual guide.}\label{fig:teaser}\end{figure}
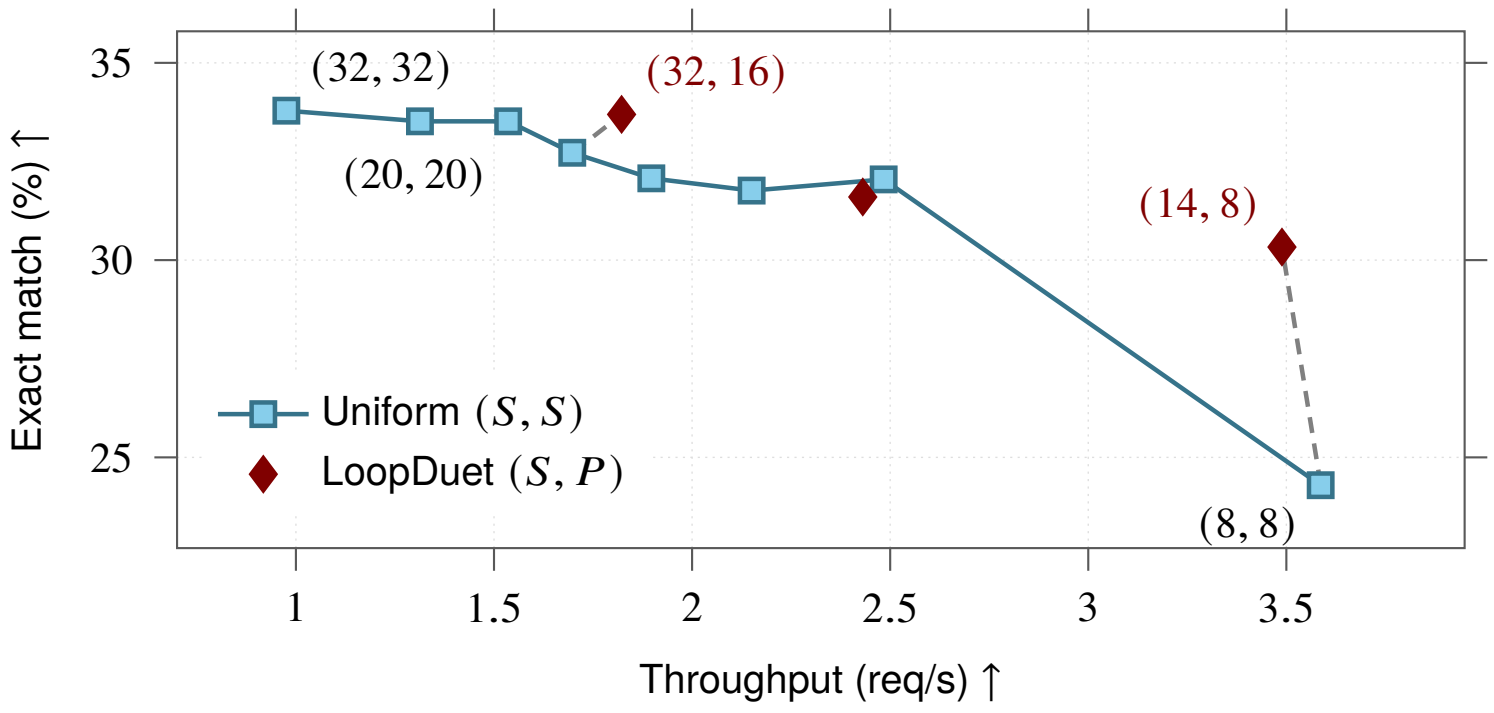
We compare uniform $(S,S)$ and LoopDuet's split-depth $(S,P)$ settings in the same serving engine. Including reduced
uniform depths tests whether separate phase budgets improve the quality--throughput choices
available to deployment. All evaluated configurations use the complete HotpotQA dataset, $C=64$, and
fixed within-model prompts, initialization, KV layout, precision, and execution mode. Throughput is
completed requests divided by wall time. The appendix lists all 11 Huginn and six OLMo
configurations with EM and F1.

\paragraph{Huginn retains answer quality at lower prefilling cost.}
Separate budgets benefit Huginn at both higher-quality and higher-throughput operating points (Figure~\ref{fig:teaser}). At the
higher-quality point, $(32,16)$ reaches EM 33.69 and F1 44.29 at 1.82 requests/s. The fastest tested
uniform setting meeting the same full-depth criterion is $(20,20)$, with EM 33.52 at 1.54
requests/s. Relative to that alternative, LoopDuet provides $1.19\times$ throughput and a
$+0.18$-point EM difference. Relative to full $(32,32)$, it provides $1.87\times$ throughput with a
$-0.09$-point EM change.

The benefit also extends to higher-throughput settings. At $(14,8)$, Huginn reaches 30.33 EM at 3.49
requests/s, compared with 24.29 EM at 3.59 requests/s for uniform $(8,8)$. Retaining deeper decoding
therefore adds 6.04 EM and 7.13 F1 points while preserving 97\% of uniform-eight throughput. This
point trades quality relative to full depth for capacity, while retaining more quality than uniform
reduction at nearby cost.

\begin{table}[t]
\centering\small
\caption{For Huginn, LoopDuet $(14,8)$ adds 6.04 EM points over uniform depth eight while retaining 97\% of its throughput. Comparisons use full-set HotpotQA, $C=64$. Throughput ratios report measured LoopDuet/uniform performance for configurations whose throughputs differ by at most 10\%. Positive quality differences favor LoopDuet.}
\begin{tabular}{@{}lllrrr@{}}
\toprule
Model & LoopDuet $(S,P)$ & Uniform & Throughput ratio & $\Delta$EM & $\Delta$F1\\
\midrule
Huginn & $(14,8)$ & $(8,8)$ & 0.973 & +6.04 & +7.13\\
Huginn & $(24,12)$ & $(12,12)$ & 0.979 & -0.45 & -0.47\\
Huginn & $(32,16)$ & $(16,16)$ & 0.960 & +1.62 & +1.90\\
Huginn & $(32,16)$ & $(18,18)$ & 1.073 & +0.97 & +1.10\\
OLMo r32 & $(11,4)$ & $(4,4)$ & 0.918 & +0.32 & +0.49\\
\bottomrule
\end{tabular}

\label{tab:frontier}
\end{table}

Table~\ref{tab:frontier} compares nearby measured alternatives. Against uniform $(18,18)$, $(32,16)$
gives 7.3\% higher throughput and 0.97 more EM points; against $(16,16)$, it gives 1.62 more EM
points at 96\% of throughput.
The results also show why prefilling and decoding depths must be selected jointly. At $(24,12)$, Huginn gives
31.60 EM at 2.43 requests/s, while uniform $(12,12)$ gives 32.05 EM at 2.48 requests/s, an EM
difference of $-0.45$. Additional decoding loops therefore benefit specific joint allocations.

\paragraph{OLMo selects a uniform starting point.}
For OLMo HotpotQA, uniform $(4,4)$ reaches 56.46 requests/s with EM 37.95, a $-0.54$-point change
from $(32,32)$. Across the six configurations, EM ranges from 37.95 to 38.49;
$(4,4)$ provides the highest throughput. LoopDuet $(11,4)$ gives 51.85 requests/s and EM 38.27, a $+0.32$-point
difference from uniform four, with F1 50.12 versus 49.63. The preferred point depends on the
deployment's quality objective; under the stated EM criterion, uniform four supplies the largest
measured throughput.

Both models produce short HotpotQA answers, yet their accuracy--throughput trade-offs favor different depth allocations. Joint
selection therefore determines whether a workload benefits from deeper decoding at a given
prefilling budget. It expands Huginn's options through split budgets and selects OLMo's uniform
$(4,4)$ point as the starting budget for load control (Section~\ref{sec:adaptive}).

\subsection{Prefilling-Loop and Batched Speedups}
\label{sec:serving}

\paragraph{A layer-pass model bounds the savings.}
Split execution assigns $c_P$ to the first $L-1$ prompt positions and $c_S$ to
the final prompt position and all decoding positions. The ratio of uniform to
split work is
\begin{equation}
  R_{\mathrm{layer}} = \frac{(L+D)\,c_S}{(L-1)\,c_P + (D+1)\,c_S}.
  \label{eq:layer}  
\end{equation}
With $M$ prompt chunks, replacing $L-1$ and $D+1$ in the denominator by $L-M$
and $D+M$ accounts for the full-depth chunk endpoints. For long prompts with
short answers, the ratio approaches $c_S/c_P$: 3.57 for OLMo $(32,8)$, 2.31
for OLMo $(11,4)$, and 1.94 for Huginn $(32,16)$. The model omits adapter,
vocabulary-head, memory-traffic, and scheduling costs.

\paragraph{Fixed decoding depth isolates the prefilling savings.}
\begin{figure}[t]\centering
\begin{tikzpicture}
\begin{groupplot}[group style={group size=3 by 2,horizontal sep=.85cm,vertical sep=1.25cm},paperaxis,width=.285\textwidth,height=2.9cm,ymin=0,ybar,/pgf/bar width=13pt,enlarge x limits=.65,xtick={0,1},grid=major,axis on top=false,nodes near coords,nodes near coords style={font=\plotfont,text=figureInk},every node near coord/.append style={/pgf/number format/fixed,/pgf/number format/precision=2}]
\nextgroupplot[ymax=36.056243052891396,xticklabels={{$(32,32)$},{$(32,8)$}},ylabel={Throughput (req/s) $\uparrow$},legend to name=servinglegend,legend columns=1,legend style={row sep=4pt}]
\addplot[fullbar,bar shift=0pt] coordinates {(0,9.156204123631067)};
\addlegendentry{Uniform $(S,S)$}
\addplot[splitbar,bar shift=0pt] coordinates {(1,27.315335646129842)};
\addlegendentry{LoopDuet $(S,P)$}
\nextgroupplot[ymax=68.44259894230159,xticklabels={{$(11,11)$},{$(11,4)$}},ylabel={}]
\addplot[fullbar,bar shift=0pt] coordinates {(0,24.30197516162438)};
\addplot[splitbar,bar shift=0pt] coordinates {(1,51.850453744167865)};
\nextgroupplot[ymax=2.4050965083282594,xticklabels={{$(32,32)$},{$(32,16)$}},ylabel={}]
\addplot[fullbar,bar shift=0pt] coordinates {(0,0.9758979671016994)};
\addplot[splitbar,bar shift=0pt] coordinates {(1,1.8220428093395904)};
\nextgroupplot[ymax=4.392335845917114,xticklabels={{$(32,32)$},{$(32,8)$}},ylabel={Median TTFT (s) $\downarrow$}]
\addplot[fullbar,bar shift=0pt] coordinates {(0,3.3275271559978137)};
\addplot[splitbar,bar shift=0pt] coordinates {(1,1.1075738260042272)};
\nextgroupplot[ymax=1.656467654630542,xticklabels={{$(11,11)$},{$(11,4)$}},ylabel={}]
\addplot[fullbar,bar shift=0pt] coordinates {(0,1.254899738356471)};
\addplot[splitbar,bar shift=0pt] coordinates {(1,0.5844002560006629)};
\nextgroupplot[ymax=68.55238042620651,xticklabels={{$(32,32)$},{$(32,16)$}},ylabel={}]
\addplot[fullbar,bar shift=0pt] coordinates {(0,51.93362153500493)};
\addplot[splitbar,bar shift=0pt] coordinates {(1,27.779241043986985)};
\end{groupplot}
\node[fit=(group c1r1.north west)(group c3r2.south east),inner sep=0pt] (servingplots) {};
\node[anchor=west] at ([xshift=3mm]servingplots.east) {\pgfplotslegendfromname{servinglegend}};
\node[anchor=north,font=\plotfont\bfseries] at ([yshift=-7mm]group c1r2.south) {(a) OLMo r32: $S=32$};
\node[anchor=north,font=\plotfont\bfseries] at ([yshift=-7mm]group c2r2.south) {(b) OLMo r32: $S=11$};
\node[anchor=north,font=\plotfont\bfseries] at ([yshift=-7mm]group c3r2.south) {(c) Huginn: $S=32$};
\end{tikzpicture}
\caption{\textbf{Removing prefilling loops increases throughput and reduces time to first token.} Full-set HotpotQA, $C=64$, one A100-80GB. Each column compares the same model and decoding depth: OLMo uses four paragraphs and Huginn ten. Top: request throughput. Bottom: median closed-loop TTFT, including scheduling effects. Each axis uses absolute units and its own scale. The throughput ratios are $2.98\times$, $2.13\times$, and $1.87\times$; the corresponding EM changes are $-0.07$, $-0.16$, and $-0.09$ points.}
\label{fig:serving}
\end{figure}
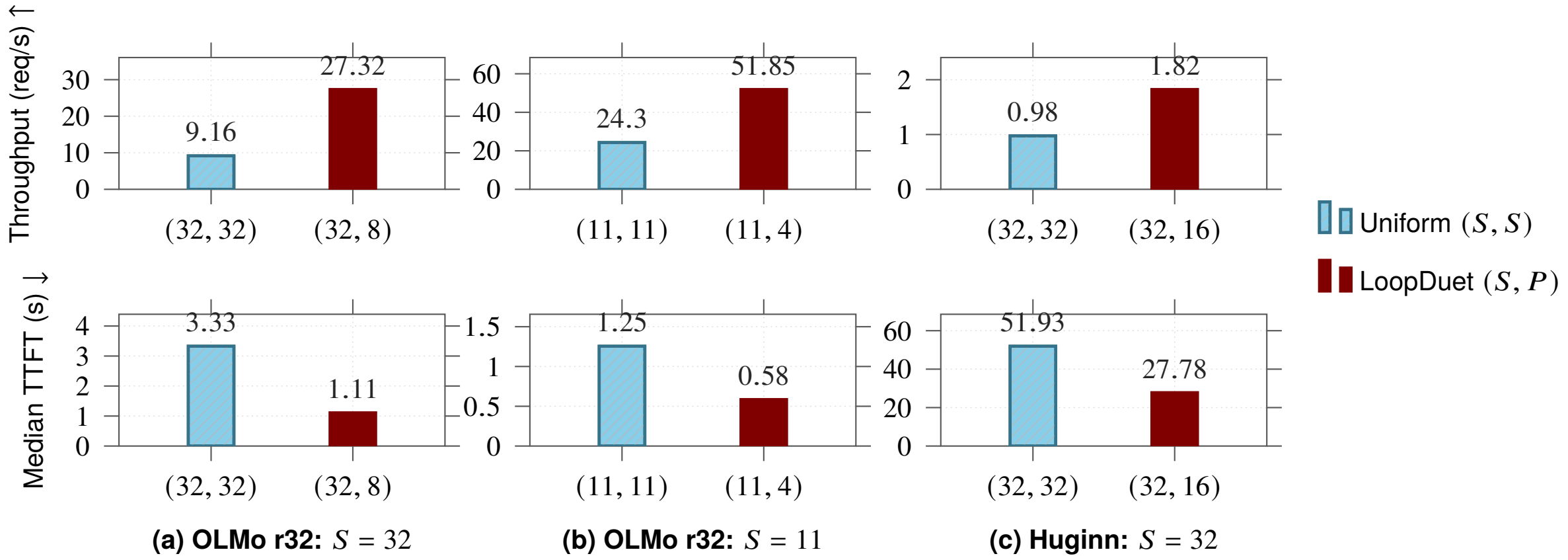
To isolate prefilling savings, we hold decoding depth fixed. Figure~\ref{fig:serving} shows OLMo's
808.7-second run falling to 271.1 seconds and Huginn's 7,588-second run to 4,064 seconds. The corresponding EM changes are $-0.07$, $-0.16$, and $-0.09$ points. These fixed-$S$ ablations isolate prefilling savings; joint depth comparisons guide the choice of decoding depth.

For the three $C=64$ configurations, measured speedups reach 85--96\% of the layer-pass ratios,
consistent with compact dense execution realizing most of the arithmetic savings. These are
within-model ratios; absolute OLMo and Huginn times also reflect different widths and context
constructions.

\paragraph{The savings persist at serving concurrency.}
Concurrency and token budgets constrain different aspects of the serving batch. We vary both to test
whether the savings persist as the executed grouping changes. Huginn $(32,16)$ sustains $1.87\times$
throughput across $C=32,64,128$ at the default 8,192-token engine-step budget, relative to matched
$(32,32)$ controls.

At $C=128$, token budgets of 4,096, 8,192, and 16,384 yield speedups of $1.84\times$, $1.87\times$,
and $1.88\times$, with EM differences of $+0.12$, $-0.09$, and $-0.14$ points. The reduced-depth
runs execute 3,797, 1,902, and 957 engine steps, respectively. This sweep changes the executed
batching regime while retaining the prefilling-depth benefit. Median TTFT falls from 126.6 to 68.7
seconds at budget 4,096 and from 103.1 to 54.8 seconds at 16,384.
OLMo shows the same consistency across $C=32,64,128$: $(11,4)$ gives speedups of $2.10\times$,
$2.13\times$, and $2.09\times$; $(32,8)$ gives $2.92\times$, $2.98\times$, and $2.98\times$.

\paragraph{Less prefilling work also reduces interference.}
Prefilling and decoding share the engine schedule, so removing prefilling work can improve latency
in both phases. At OLMo $(11,4)$, median closed-loop TTFT falls from 1.26 to 0.58 seconds and median
TPOT from 0.420 to 0.196 seconds. The share with TTFT at most one second rises from 44.2\% to
99.7\%. Decoding retains the same loops per token; the TPOT reduction accompanies the lighter
prefilling workload in the schedule. At $C=128$, measured request throughput rises from 24.8 to 51.8
requests/s. Huginn's median TTFT falls from 51.9 to 27.8 seconds.

\textbf{Workload and granularity determine the payoff.}
The payoff depends on workload composition. HotpotQA's long prompts and short answers yield the
measured throughput gains, whereas GSM8K's longer responses give wall-time ratios of 0.98--1.01 across four serving pairs despite layer-pass ratios of 1.52--2.71. 
Amdahl's law explains this gap:
\begin{equation}
  R_{\mathrm{time}} \approx \bigl[(1-f) + f/r + \epsilon\bigr]^{-1},
  \label{eq:amdahl}  
\end{equation}
where $f$ is the baseline wall-time fraction spent on the affected work, $r$ its acceleration, and $\epsilon$ the added overhead normalized by baseline time. Longer responses lower $f$, which bounds the gain by $1/(1-f)$ regardless of $r$. The appendix reports the GSM8K timings.

Selection granularity also determines whether savings survive batching. In the FFN-routing study,
retaining half the neurons per token gives some accuracy-preserving configurations, yet their union
covers 99.80--99.95\% of neurons at batch size 16. A common prefilling-loop boundary instead removes
later core applications from ordinary prompt positions across the batch. Complete routing and union results appear in the appendix.

\subsection{Queue Control Retains Quality under Load}
\label{sec:adaptive}
\begin{table}[t]
\centering\small
\caption{One-second TTFT attainment (\%) / EM on all HotpotQA examples, OLMo r32, $S=4$, four-paragraph construction, open-loop $C=128$. The same queue-control parameters apply to all four traces. LoopDuet attains the one-second target for every request on the first three traces and for 87.5\% on the 20/80 burst trace. Bold highlights LoopDuet.}
\begin{tabular}{@{}lrrrr@{}}
\toprule
Arrival trace & Fixed $(4,4)$ & Fixed $(4,3)$ & Fixed $(4,2)$ & \shortstack{LoopDuet (queue)\\$(4,P\in[2,4])$}\\
\midrule
Poisson 40 req/s & 100 / 37.92 & 100 / 36.29 & 100 / 31.83 & \textbf{100 / 37.96}\\
Poisson 60 req/s & 8.6 / 38.04 & 100 / 36.35 & 100 / 31.74 & \textbf{100 / 37.27}\\
Burst 10/60 req/s & 46.9 / 38.04 & 100 / 36.42 & 100 / 31.82 & \textbf{100 / 37.77}\\
Burst 20/80 req/s & 21.3 / 38.20 & 39.1 / 36.42 & 100 / 31.91 & \textbf{87.5 / 35.12}\\
\bottomrule
\end{tabular}
\label{tab:controller}
\end{table}
\paragraph{More quality at the required capacity.}
Starting from OLMo's selected $(4,4)$ point, LoopDuet's controller keeps $S=4$ and adjusts $P\in\{2,3,4\}$. Fixed prefilling depths at the same $S$ provide the controls. All policies process the same dataset and seeded arrivals, with parameters fixed across loads.

At Poisson 60 requests/s, the controller retains 0.92 more EM points than the most accurate tested
fixed depth attaining the one-second TTFT target for every request. Fixed $P=4$ gives EM 38.04 at
8.6\% attainment; fixed $P=3$ reaches 100\% at EM 36.35. The controller also reaches 100\%, with EM
37.27 and median / 99th-percentile TTFT of 0.486 / 0.725 seconds.

The same benefit holds under alternating 10/60 requests/s demand. The controller reaches EM 37.77 at
100\% attainment, a 1.35-point gain over fixed $P=3$ at EM 36.42. Its TTFT p99 is 0.682 seconds. The
trace spends approximately 87\% of wall time at $P=4$ and 13\% at $P=3$, restoring depth when queue
age permits. These are controller-state time fractions, not token-weighted per-request depths.
\begin{figure}[t]\centering
\begin{tikzpicture}\begin{axis}[paperaxis,width=.99\ceiplotwidth,height=4.5cm,xmin=31,xmax=39.1,ymin=35,ymax=112,xlabel={Exact match (\%) $\uparrow$},ylabel={TTFT $\leq1$ s (\%) $\uparrow$},label style={font=\plotfont},tick label style={font=\plotfont},legend style={font=\plotfont,draw=none,at={(.02,.26)},anchor=west},axis lines=box]
\addplot[only marks,uniformline,mark size=2.5pt] coordinates {(38.04,46.9)(36.42,100)(31.82,100)};\addlegendentry{Fixed prefilling depth}
\addplot[splitmarks] coordinates {(37.77,100)};\addlegendentry{LoopDuet (queue)}
\draw[densely dashed,gray] (axis cs:36.42,100)--(axis cs:37.77,100);
\node[font=\plotfont,anchor=south] at (axis cs:37.1,102) {$+1.35$ EM};
\node[font=\plotfont,anchor=north east] at (axis cs:38.04,46.9) {$P=4$};
\node[font=\plotfont,anchor=north] at (axis cs:36.42,98) {$P=3$};
\node[font=\plotfont,anchor=north] at (axis cs:31.82,98) {$P=2$};
\end{axis}\end{tikzpicture}
\caption{\textbf{Adaptation retains more quality at full measured SLO attainment.} OLMo at $S=4$ on the 10/60 requests/s burst trace. LoopDuet's queue controller gains 1.35 EM points over fixed $P=3$, with zero one-second TTFT violations.}\label{fig:controller}\end{figure}
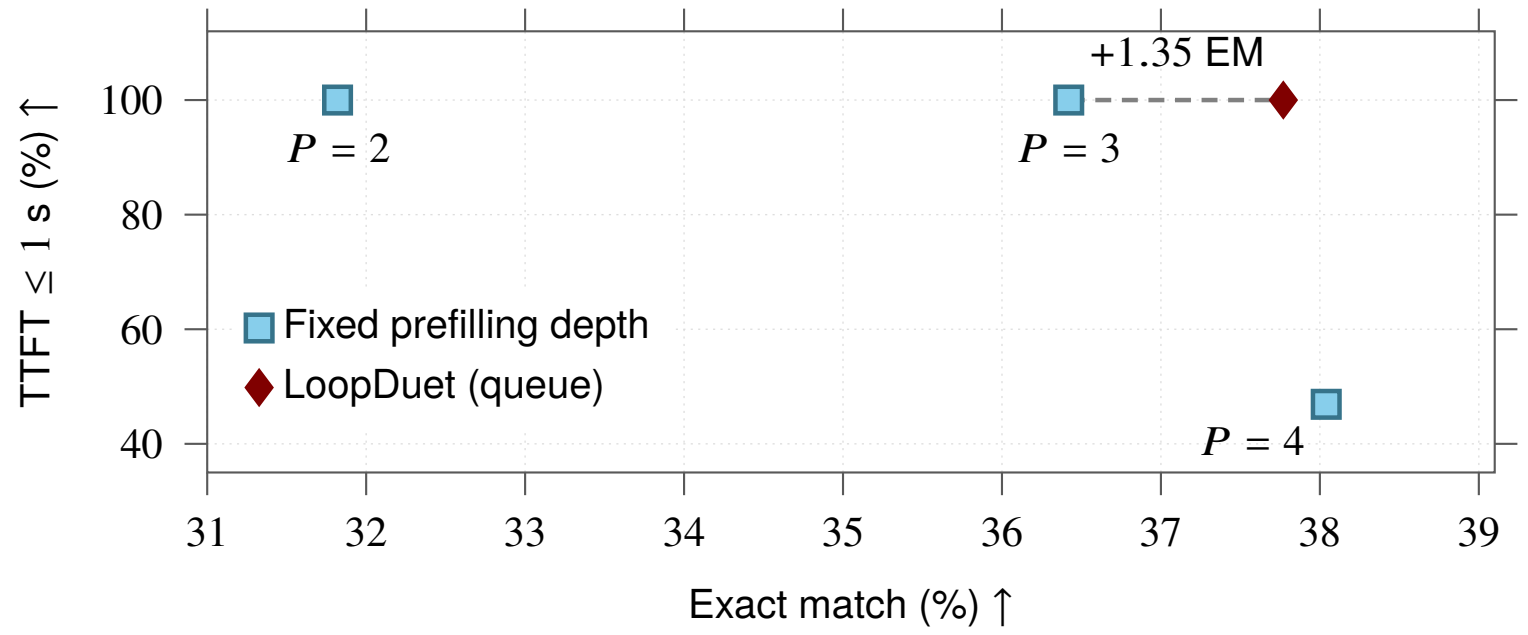
At Poisson 40 requests/s, capacity is sufficient and the controller remains at $P=4$. Both it and
fixed $P=4$ attain the target for every request, with EM 37.96 and 37.92, respectively.

\textbf{The operating envelope sets the quality--latency trade-off.}
On the stronger 20/80 burst trace, the controller gives EM 35.12 and 87.5\% attainment, with TTFT
p99 1.110 seconds. Fixed $P=2$ reaches 100\% at EM 31.91. The controller's $+3.21$-point EM
difference is therefore a quality--latency trade-off, distinct from the equal-attainment gains
above.

The appendix examines how the depth range and control signal affect queueing and retained quality.

\subsection{Token Role Explains Depth Sensitivity}
\label{sec:accuracy}

\paragraph{The first output token needs full depth.}
We first reduce only the final prompt position. For Huginn on HotpotQA with
$S = 32$ and ordinary prompt positions at $P = 16$, running the final position
at 16 instead of 32 loops lowers EM by 1.74 points relative to $(32,16)$ and
by 1.97 points relative to $(32,32)$. Keeping this position at $S$ costs one
extra full-depth position per request.

\begin{figure}[t]
\centering
\begin{tikzpicture}
\begin{axis}[
 paperaxis,width=.90\ceiplotwidth,height=4.65cm,
 xbar,bar width=6pt,
 xmin=-15.0,xmax=5.5,ymin=-.55,ymax=2.55,
 xtick={-12,-8,-4,0,4},ytick={0,1,2},
 yticklabels={{Huginn\\$(14,8)$},{OLMo r16\\$(16,4)$},{OLMo r32\\$(11,4)$}},
 yticklabel style={align=right,font=\plotfont},
 xlabel={Accuracy change vs.\ $(S,S)$ (pp) $\uparrow$},
 xmajorgrids=true,ymajorgrids=false,
 axis on top=false,clip=false,
 legend columns=2,legend style={at={(.5,1.04)},anchor=south,column sep=5pt},
 nodes near coords={\pgfmathprintnumber[fixed,fixed zerofill,precision=2,showpos]{\pgfplotspointmeta}},
 nodes near coords style={font=\plotfont,text=figureInk},
]
\draw[black!60,densely dashed,line width=.55pt] (axis cs:0,-.55)--(axis cs:0,2.55);
\addplot[fullbar,bar shift=5pt] table[x=uniform_delta,y=row,col sep=comma]{figures/role_gsm8k.csv};
\addlegendentry{Uniform $(P,P)$}
\addplot[splitbar,bar shift=-5pt] table[x=split_delta,y=row,col sep=comma]{figures/role_gsm8k.csv};
\addlegendentry{LoopDuet $(S,P)$}
\end{axis}
\end{tikzpicture}
\caption{\textbf{Deeper decoding preserves GSM8K accuracy across checkpoints.}
At the same prefilling depth, LoopDuet gains 5.91, 3.18, and 11.68 accuracy points over
uniform reduction for OLMo r32, OLMo r16, and Huginn, respectively. Bars show changes from
each checkpoint's $(S,S)$ control (zero line); row labels give $(S,P)$. LoopDuet retains
depth $S$ for decoding and the final prefilling position.}
\label{fig:role_gsm8k}
\end{figure}
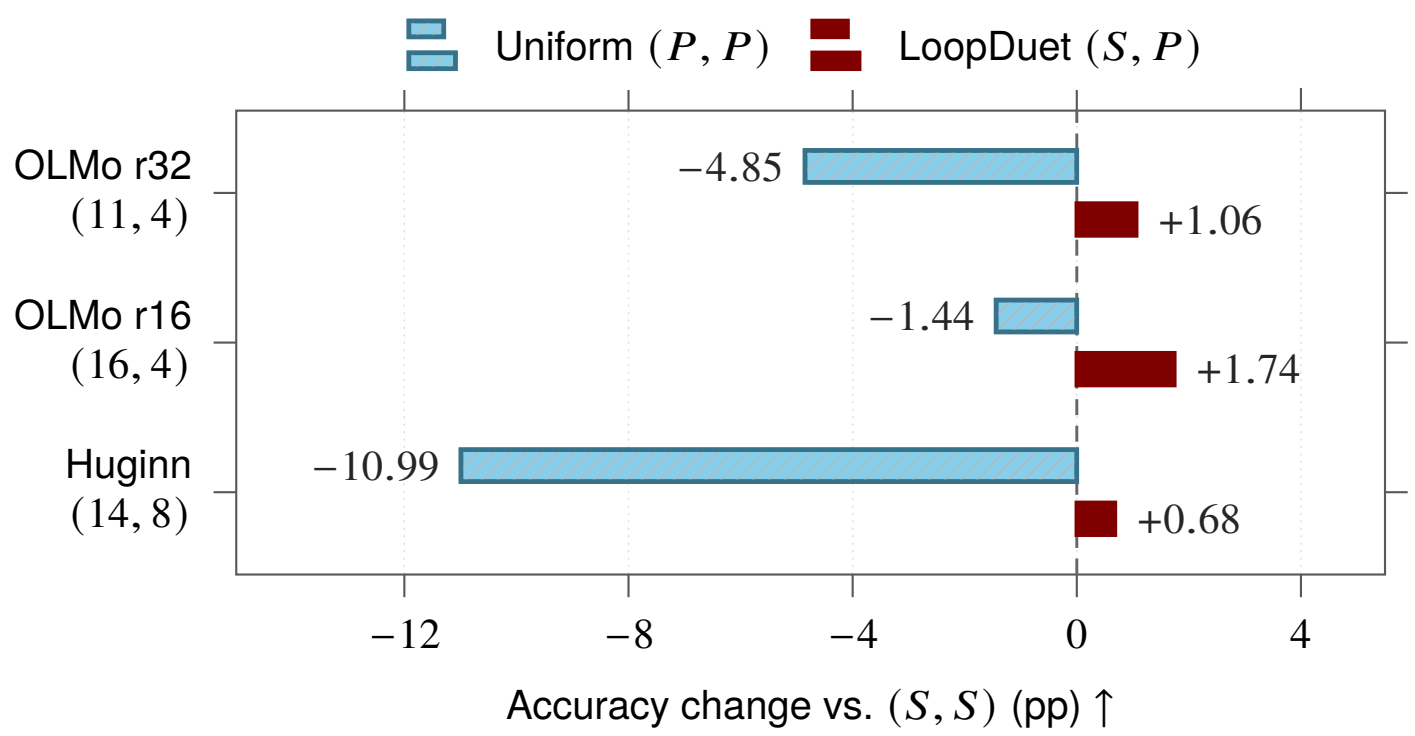

\textbf{Deeper decoding preserves GSM8K accuracy.}
At the same prefilling depth, retaining deeper decoding and first-token computation improves
GSM8K accuracy by 3.18--11.68 points across all three checkpoints
(Figure~\ref{fig:role_gsm8k}). For OLMo r32, $(11,4)$ changes accuracy by $+1.06$ points relative
to $(11,11)$, whereas uniform $(4,4)$ changes it by $-4.85$ points. Huginn $(14,8)$ changes
accuracy by $+0.68$ points relative to its depth-14 control while uniform depth eight loses
10.99 points.

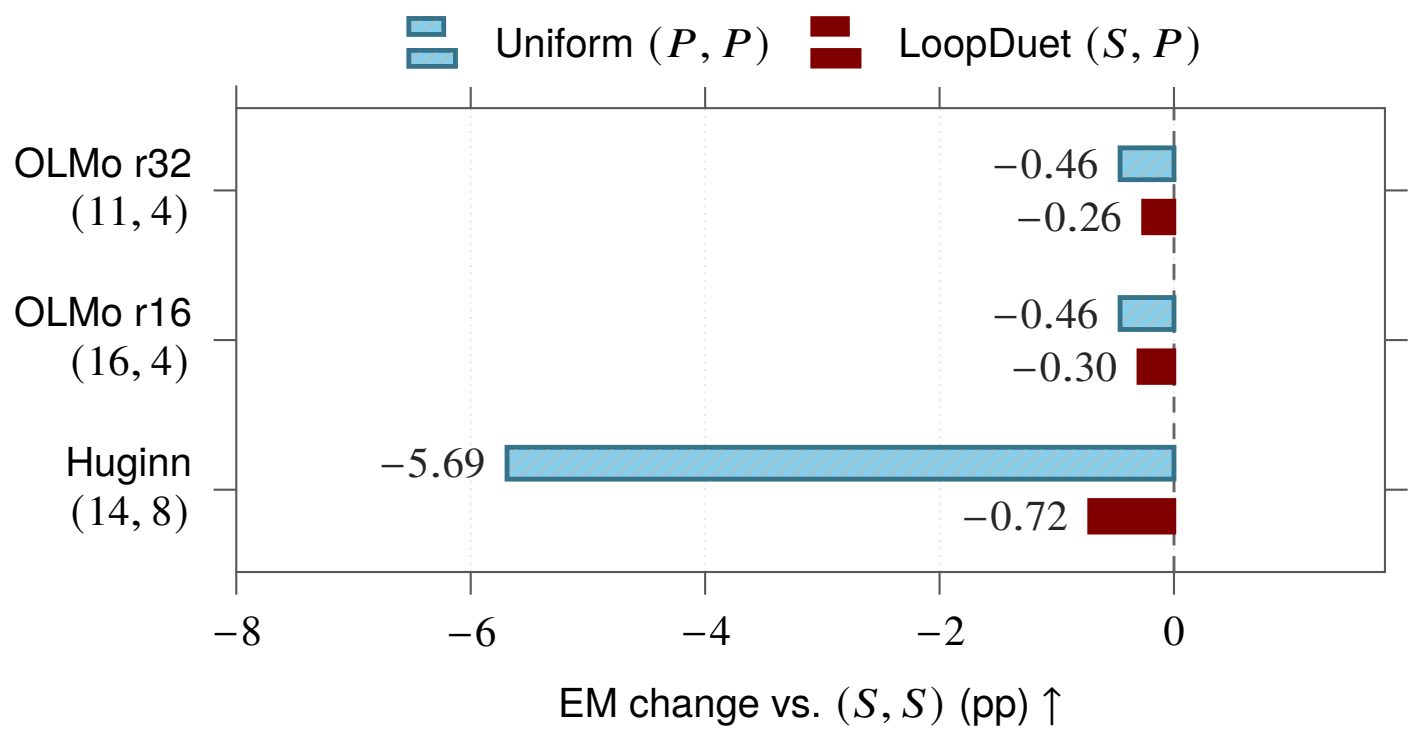
\begin{figure}[t]
\centering
\begin{tikzpicture}
\begin{axis}[
 paperaxis,width=.90\ceiplotwidth,height=4.65cm,
 xbar,bar width=6pt,
 xmin=-8.0,xmax=1.8,ymin=-.55,ymax=2.55,
 xtick={-8,-6,-4,-2,0},ytick={0,1,2},
 yticklabels={{Huginn\\$(14,8)$},{OLMo r16\\$(16,4)$},{OLMo r32\\$(11,4)$}},
 yticklabel style={align=right,font=\plotfont},
 xlabel={EM change vs.\ $(S,S)$ (pp) $\uparrow$},
 xmajorgrids=true,ymajorgrids=false,
 axis on top=false,clip=false,
 legend columns=2,legend style={at={(.5,1.04)},anchor=south,column sep=5pt},
 nodes near coords={\pgfmathprintnumber[fixed,fixed zerofill,precision=2,showpos]{\pgfplotspointmeta}},
 nodes near coords style={font=\plotfont,text=figureInk},
]
\draw[black!60,densely dashed,line width=.55pt] (axis cs:0,-.55)--(axis cs:0,2.55);
\addplot[fullbar,bar shift=5pt] table[x=uniform_delta,y=row,col sep=comma]{figures/role_hotpotqa.csv};
\addlegendentry{Uniform $(P,P)$}
\addplot[splitbar,bar shift=-5pt] table[x=split_delta,y=row,col sep=comma]{figures/role_hotpotqa.csv};
\addlegendentry{LoopDuet $(S,P)$}
\end{axis}
\end{tikzpicture}
\caption{\textbf{Huginn benefits more from deeper decoding on HotpotQA.}
LoopDuet $(14,8)$ gains 4.97 EM points over uniform depth eight and stays within 0.72 points
of its depth-14 control. The OLMo gains are 0.20 and 0.16 points. Bars show EM changes from
each $(S,S)$ control (zero line); row labels give $(S,P)$.}
\label{fig:role_hotpotqa}
\end{figure}

\textbf{The benefit depends on the model and task.}
On HotpotQA, Huginn $(14,8)$ gains 4.97 EM points over uniform depth eight and remains within
0.72 points of its depth-14 control. The OLMo comparisons change by 0.16--0.20 points at the
same prefilling depth (Figure~\ref{fig:role_hotpotqa}). This difference explains why retaining
deeper decoding is more valuable for Huginn on this task.

\begin{figure}[t]
\centering
\begin{tikzpicture}
\begin{axis}[
 paperaxis,width=.90\ceiplotwidth,height=4.65cm,
 xbar,bar width=6pt,
 xmin=-15.0,xmax=5.5,ymin=-.55,ymax=2.55,
 xtick={-12,-8,-4,0,4},ytick={0,1,2},
 yticklabels={{Huginn\\$(14,8)$},{OLMo r32\\$(11,8)$},{OLMo r32\\$(11,4)$}},
 yticklabel style={align=right,font=\plotfont},
 xlabel={Accuracy change vs.\ $(S,S)$ (pp) $\uparrow$},
 xmajorgrids=true,ymajorgrids=false,
 axis on top=false,clip=false,
 legend columns=2,legend style={at={(.5,1.04)},anchor=south,column sep=5pt},
 nodes near coords={\pgfmathprintnumber[fixed,fixed zerofill,precision=2,showpos]{\pgfplotspointmeta}},
 nodes near coords style={font=\plotfont,text=figureInk},
]
\draw[black!60,densely dashed,line width=.55pt] (axis cs:0,-.55)--(axis cs:0,2.55);
\addplot[fullbar,bar shift=5pt] table[x=uniform_delta,y=row,col sep=comma]{figures/role_arc.csv};
\addlegendentry{Uniform $(P,P)$}
\addplot[splitbar,bar shift=-5pt] table[x=split_delta,y=row,col sep=comma]{figures/role_arc.csv};
\addlegendentry{LoopDuet $(S,P)$}
\end{axis}
\end{tikzpicture}
\caption{\textbf{ARC also requires sufficient prefilling depth.}
At $S=11$, increasing OLMo prefilling depth from four to eight changes the accuracy delta
from $-1.54$ to $+0.85$ points. Huginn $(14,8)$ gains 5.63 points over uniform depth eight,
with a $-4.44$-point change from $(14,14)$. Bars use the same-$S$ control as zero;
row labels give $(S,P)$.}
\label{fig:role_arc}
\end{figure}
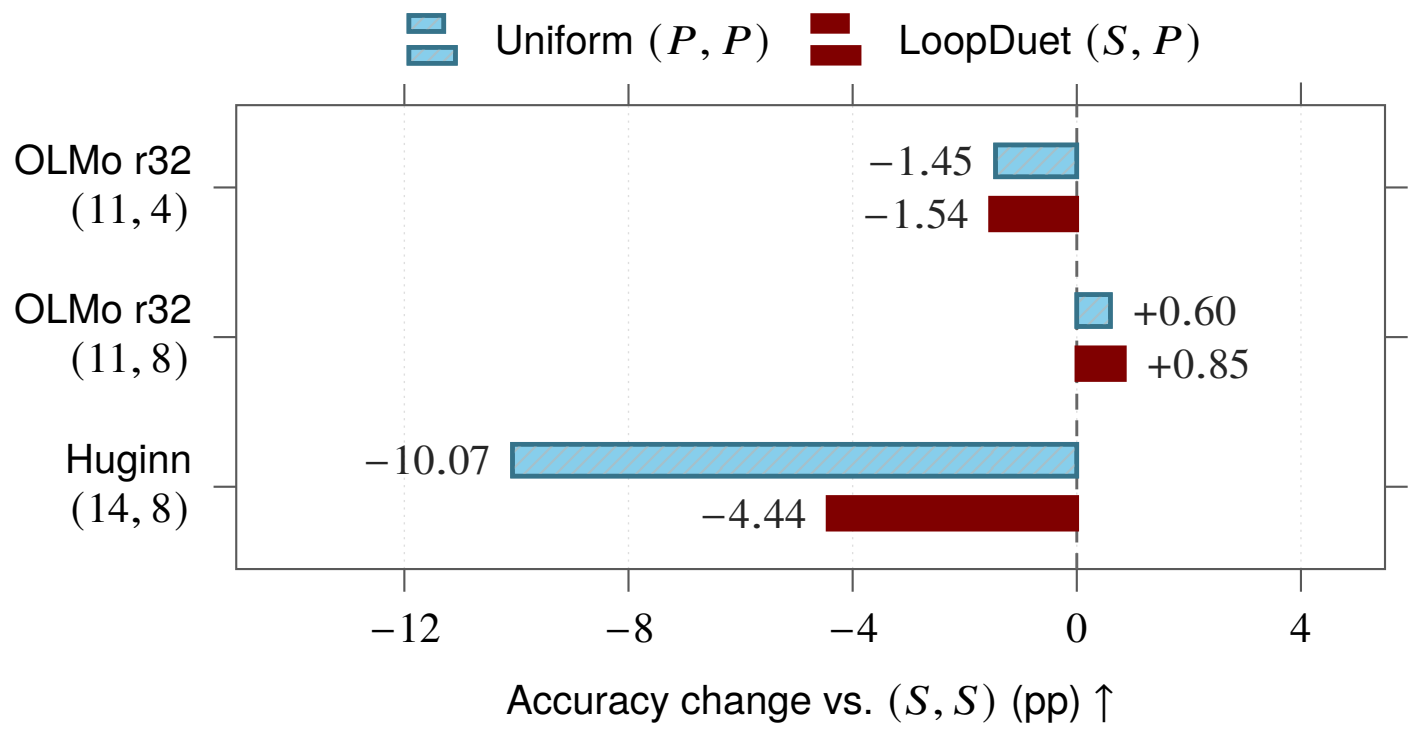

\textbf{Prefilling depth must also match task sensitivity.}
On ARC, increasing OLMo r32's prefilling depth from four to eight at $S=11$ changes the
accuracy delta from $-1.54$ to $+0.85$ points (Figure~\ref{fig:role_arc}).
Huginn $(14,8)$ gains 5.63 points over uniform depth eight but loses 4.44 points relative to
$(14,14)$. Deeper decoding improves this allocation, while the remaining accuracy cost motivates
task-specific prefilling depth.

\textbf{The serving cache retains the effect.}
Shared-KV harness comparisons yield deltas of $-0.23$ and $-0.09$ for OLMo GSM8K and HotpotQA, and
$-0.38$ and $-0.49$ for Huginn. The Huginn HotpotQA comparison is $(32,16)$ versus $(32,32)$.

\subsection{Recurrent States Prefilling Depths}
\label{sec:distance}
Role-specific state changes help explain depth sensitivity and identify candidate prefilling depths.
Distances are measured separately for ordinary prompt tokens, the final prefilling token, and
decoding tokens. Replaying recorded answers through $S_{\mathrm{ref}}$ loops supplies the states at each intermediate loop. For token $i$ and a candidate decoding depth $S\leq
S_{\mathrm{ref}}$, define
\begin{equation}
 d_i(P;S)=\frac{\lVert h_i^P-h_i^S\rVert_2}{\lVert h_i^S\rVert_2}.
\end{equation}
Both states come from the same trajectory. Curves to the terminal reference use $S=S_{\mathrm{ref}}$; lower-$S$ diagnostic cells use the corresponding pairwise intermediate-loop distances. The available traversals use $S_{\mathrm{ref}}=32$ for OLMo r32 and Huginn and 16 for OLMo r16. These are finite-depth reference states, rather than demonstrated mathematical fixed points.

At four loops, OLMo r32 GSM8K has prefilling-token and decoding-token 90th-percentile distances 0.18
and 0.20. Prefilling reduction to $(32,4)$ leaves measured accuracy at 48.29, equal to $(32,32)$.
Uniform reduction to $(4,4)$ gives 42.53 ($-5.76$ points). Thus comparable distance magnitudes
accompany different quality outcomes depending on the role receiving reduced depth. Since uniform
reduction changes both prefilling and decoding computation, this is evidence of role sensitivity
rather than an isolated causal intervention on decoding states.

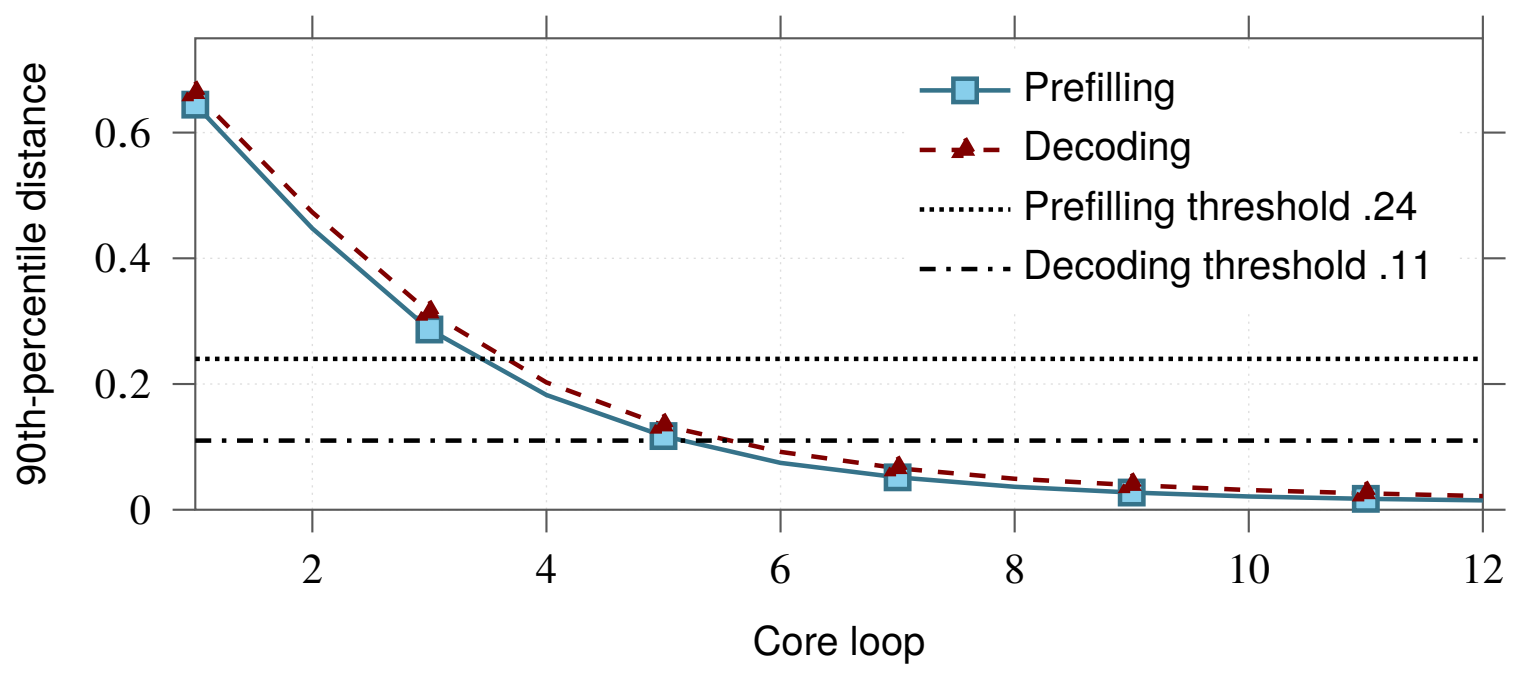
\begin{figure}[t]\centering
\begin{tikzpicture}\begin{axis}[paperaxis,width=.99\ceiplotwidth,height=4.7cm,xmin=1,xmax=12,ymin=0,ymax=.75,xlabel={Core loop},ylabel={90th-percentile distance},label style={font=\plotfont},tick label style={font=\plotfont},legend style={font=\plotfont,draw=none,at={(.98,.98)},anchor=north east},axis lines=box]
\addplot[uniformline,mark repeat=2] table[x=loop,y=prompt,col sep=comma]{figures/convergence.csv};\addlegendentry{Prefilling}
\addplot[methodMaroon,dashed,mark=triangle*,mark repeat=2] table[x=loop,y=answer,col sep=comma]{figures/convergence.csv};\addlegendentry{Decoding}
\addplot[black,densely dotted] coordinates {(1,.24)(12,.24)};\addlegendentry{Prefilling threshold .24}
\addplot[black,dashdotted] coordinates {(1,.11)(12,.11)};\addlegendentry{Decoding threshold .11}
\end{axis}\end{tikzpicture}
\caption{\textbf{State distance has role-dependent quality implications.} OLMo r32 GSM8K, depth-32 reference. At loop four, ordinary prompt tokens (prefilling) and generated tokens (decoding) have distances 0.18 and 0.20, but their depth interventions have different accuracy effects. Horizontal lines show thresholds fitted across the evaluated depth configurations.}\label{fig:convergence}\end{figure}

Across the measured cells, the best separating threshold on the prompt 90th-percentile distance is
approximately 0.24. Applying the same procedure to decoding-token distances for uniform-depth cells
gives approximately 0.11. These fits associate larger prompt-state deviations with preserved quality
under prefilling reduction. The roughly twofold difference describes the evaluated cells and roles;
it does not require prompt states always to settle earlier. The supplement reports the
crossing-depth comparisons.

\textbf{One reference trace supplies candidate statistics.}
Let $q_{.9}(P;S)$ be the 90th percentile of the ordinary-prompt distances for a candidate $(S,P)$. We propose candidate depths satisfying
\begin{equation}
 q_{.9}(P;S)\leq\tau,\qquad \tau\approx0.24.
\end{equation}
One replay of recorded answers supplies state statistics for all candidate loop pairs, including lower-$S$ cells extracted from the same trajectory. We then evaluate the proposed depths on task accuracy.

Fitting on two checkpoints and evaluating the third gives thresholds between 0.237 and 0.241,
close to the pooled value of 0.24. The distance curves suggest candidate prefilling depths around
3--4 for OLMo and 13--16 for Huginn. The appendix compares the resulting selections with
task-quality measurements and alternative selection rules.

Task sensitivity matters even at small state distances. On ARC, OLMo $(32,4)$ and $(11,4)$
lose 1.88 and 1.54 accuracy points, and Huginn $(32,16)$ loses 1.45 points, all at distances
below 0.18. OLMo GSM8K $(32,8)$ loses 1.29 points at distance 0.037. Conversely, Huginn GSM8K
$(14,8)$ gains 0.68 points at distance 0.614. These observations suggest that the task relevance
of a state change matters alongside its magnitude, motivating task-specific depth selection.
The appendix examines these patterns, aggregation choices, and prompt composition.

\section{Related Work}
\textbf{Phase-specific prefilling computation.}
Prior methods reduce prefilling cost through approximation or architectural separation. SwiftKV
distills later-layer prompt cache states \citep{qiao-etal-2025-swiftkv}, while POP skips deep prefilling and
preserves KV projections and first-token computation \citep{he2026popprefillonlypruningefficient}. SPEED changes prompt visibility
using representation diagnostics \citep{oh2026shallowprefilldeepdecoding}. Decode-Branch Transformers separate prefilling/cache
and decoding paths \citep{liu2026decodebranchtransformersdecouplingprimary}, and YOCO enables prefilling exit through a shared self-decoder
cache \citep{sun2024you}. LoopDuet connects joint depth selection on existing recurrent checkpoints to batched
execution and queue-driven prefilling budgets.

\textbf{Parameter sharing and recurrent depth.}
Transformer blocks \citep{NIPS2017_3f5ee243} support depth-wise sharing in ALBERT, iterative refinement in
Universal Transformers, and implicit equilibrium in DEQ \citep{Lan2020ALBERT:,dehghani2018universal,NEURIPS2019_01386bd6}. Huginn and
recurrent retrofits expose repeated computation as an inference budget \citep{geiping2025scaling,mcleish2026teaching}.
Fixed-point shaping supports terminal KV sharing and distilled or shallow teacher prefilling
\citep{huang2026loopedmodelsrightii}. Other recurrent methods allocate precision (RAMP), change retained context (HLA
and shared-memory models), or train across budgets (Elastic Latent Recurrence)
\citep{wan2026ramprecurrentadaptivemixed,chen2026hybridlatentattentionlooped,monea2026surprisingeffectivenesssharedmemory,ladhak2026modelloopallelastic}. Our focus is phase-level depth allocation with existing weights.

\paragraph{Early-exiting and skip layers}
Adaptive computation selects work at several granularities. ACT and PonderNet learn halting
\citep{graves2017adaptivecomputationtimerecurrent,banino2021pondernet}; BranchyNet and SkipNet add exits or block skipping
\citep{teerapittayanon2016branchynet,Wang_2018_ECCV}. DeeBERT, FastBERT, and patience-based exits use prediction confidence or
consistency \citep{xin-etal-2020-deebert,liu-etal-2020-fastbert,NEURIPS2020_d4dd111a}. LayerDrop trains removable layers, Mixture-of-Depths
learns capacity-constrained token routing, and Mixture-of-Recursions learns token-level recurrence
through shared blocks \citep{Fan2020Reducing,raposo2024mixtureofdepthsdynamicallyallocatingcompute,bae2025mixtureofrecursions}. We select phase-level budgets that remove
batched work at a common loop boundary.

Variable-depth autoregressive inference must preserve cache consistency and efficient batching.
CALM, FREE, SkipDecode, and EE-LLM address these requirements for early exits
\citep{schuster2022confident,bae2023fast,delcorro2023skipdecodeautoregressiveskipdecoding,chen2024eellm}. Draft \& Verify and LayerSkip verify reduced-depth drafts
\citep{zhang-etal-2024-draft,elhoushi-etal-2024-layerskip}; speculative sampling preserves the target distribution through
verification \citep{pmlr-v202-leviathan23a,chen2023acceleratinglargelanguagemodel}. SpecInfer organizes token trees, Medusa adds prediction heads, and EAGLE variants construct
feature-based drafts \citep{miao2024specinfer,cai2024medusa,li2024eagle,li-etal-2024-eagle}. Looped-model systems
manage heterogeneous exits, batch recurrence steps, or speculate across position and depth
\citep{mishra2026enablingdynamiccomputationlooped,schwethelm2026depthadaptiveinferenceloopedlanguage,ha2026wavefrontdecodingparallelizedselfspeculative}. LoopDuet directly uses the retained shallower prompt states during decoding.

\textbf{Scheduling and latency control.}
Clipper, Clockwork, and AlpaServe optimize batching, predictability, and resource sharing
\citep{201468,258862,288582}; Orca and FastServe schedule iterations and preemption
\citep{280922,wu2024fastdistributedinferenceserving}. SARATHI variants mix prompt chunks with decoding, while DistServe and
Splitwise separate phase resources \citep{agrawal2023sarathiefficientllminference,298679,298687,10609649}. Parrot uses
application dependencies \citep{298764}, and Apparate adapts exits with accuracy feedback
\citep{dai2024apparate}. LoopDuet's controller adjusts prefilling work using queue age within an
offline-characterized quality--capacity range.

\textbf{Conditional computation under batching.}
MoE architectures select subsets of FFN computation per token
\citep{shazeer2017,JMLR:v23:21-0998,lepikhin2021gshard,dai-etal-2024-deepseekmoe,zhang-etal-2022-moefication}. FastMoE and MegaBlocks optimize sparse training;
Tutel and DeepSpeed-MoE optimize expert execution and parallelism
\citep{he2021fastmoefastmixtureofexperttraining,MLSYS2023_5a54f793,MLSYS2023_5616d34c,pmlr-v162-rajbhandari22a}. Shared weights also create batching opportunities in
adapter serving, as developed in LoRA, S-LoRA, and Punica \citep{hu2022lora,MLSYS2024_906419cd,MLSYS2024_054de805}. These systems
motivate considering aggregate executed work. Section~\ref{sec:discussion} develops the connection
to MoE serving.

\section{Discussion}
\label{sec:discussion}

\textbf{Recurrence as a third axis of conditional computation.}
LoopDuet's phase-dependent depth connects naturally to MoE: physical layer, expert identity,
and recurrent iteration form three axes of computation allocation, a \emph{3D MoE} abstraction.
Conventional MoE routes tokens among experts within a layer, providing the horizontal axis
\citep{shazeer2017,JMLR:v23:21-0998}. MoD and layer-skipping methods adapt computation across
layers, providing the vertical axis
\citep{raposo2024mixtureofdepthsdynamicallyallocatingcompute,elhoushi-etal-2024-layerskip}.
Recurrence adds a third axis by revisiting shared blocks: states evolve across iterations
while parameters remain tied.

\textbf{MoE principles extend to recurrent computation.}
Several developments in looped models can be understood through MoE principles.
\emph{Sparse experts expand capacity}: routing within shared blocks increases their
representational capacity, as explored by the Sparse Universal Transformer \citep{tan2023sparse},
MoEUT \citep{csordas2024moeut}, and scaling studies of looped MoE models under matched compute,
parameter, and KV budgets \citep{wang2026smelt}.
\emph{Sparse work requires efficient batching}: MoE systems group tokens assigned to the
same expert, while continuous depth batching groups tokens at different recurrence steps
for execution through a shared core
\citep{bae2025mixtureofrecursions,schwethelm2026depthadaptiveinferenceloopedlanguage}.
\emph{Prediction enables work to begin earlier}: expert prefetching has an analogue in
wavefront decoding, which uses an intermediate recurrent state to draft the next token
\citep{ha2026wavefrontdecodingparallelizedselfspeculative}.
These connections motivate extending further MoE serving techniques to recurrent models.

\textbf{Phase and batching determine the benefit of sparsity.}
Prefilling and decoding have different bottlenecks, motivating phase-specific serving
policies \citep{298687,10609649}. For recurrent models, this distinction also determines
how much depth each phase should receive. MoE serving further shows that per-token
sparsity must reduce aggregate executed work to produce system-level savings.
Lynx consolidates low-affinity token-to-expert assignments onto a smaller batch-wide
expert set to reduce memory traffic \citep{gupta2026lynxenablingefficientmoe}, while
DeepSpeed-MoE coordinates expert parallelism with inference kernels
\citep{pmlr-v162-rajbhandari22a}.
Our neuron-union measurements reveal the same aggregation effect within looped blocks:
per-token selections cover nearly every neuron across a batch. A common prefilling-depth
boundary instead removes later core applications from ordinary prefilling positions.
Lynx favors expert reassignment during decoding, while our measurements favor recurrence
reduction during prefilling.

\textbf{Joint scheduling of residency and recurrence.}
MoE serving coordinates computation with parameter movement. SiDA-MoE predicts expert
activation ahead of execution and overlaps prediction with inference to manage GPU/host
residency \citep{MLSYS2024_698cfaf7}. MoE-Infinity uses request-level activation traces
for expert caching and prefetching in memory-constrained, single-user serving
\citep{xue2025moeinfinityefficientmoeinference}.
Both schedule data movement around active experts, while depth control determines how
often a resident block is reused. These decisions interact in a recurrent MoE:
loop depth can change routing locality and prefetch opportunities, while expert residency
affects the cost of each additional iteration. This coupling motivates joint scheduling
of expert residency and per-phase recurrent depth.

\section{Conclusion}
LoopDuet jointly selects prefilling and decoding depth to expand serving choices for looped language
models. For Huginn, LoopDuet delivers 19\% more throughput than the fastest tested uniform
configuration meeting our accuracy criterion, and 6.04 more EM points than uniform depth eight at
97\% of its throughput. For OLMo HotpotQA, the accuracy–throughput comparison favors uniform depth four;
adapting its prefilling budget retains 0.92--1.35 more EM points than the most accurate tested fixed
settings achieving the same one-second TTFT attainment on the evaluated 60 requests/s and 10/60
burst traces. Role-separated measurements and batch-compatible execution connect these depth
decisions to concrete quality--throughput trade-off.

\clearpage
\subsection*{AI Usage Disclosure}
OpenAI Codex and Anthropic Claude Code were used as AI-assisted tools throughout the research and manuscript preparation process. Their contributions included implementing experimental code according to the authors’ instructions, executing experiments, organizing and refactoring code, inspecting experimental artifacts, and assisting with debugging and validation. These tools were also used to support literature searches, identify relevant prior work, and verify bibliographic references. During manuscript preparation, AI assistance was used to draft and revise text, improve clarity and presentation, generate figures and supporting validation scripts, and check the consistency of reported statistics against experimental results.
All research ideas, scientific hypotheses, methodological decisions, and experimental designs originated from the authors. The AI tools primarily served to implement the authors’ instructions and assist with technical execution and documentation. The authors reviewed and verified the core implementations, experimental results, and scientific claims, and retained full responsibility for the accuracy, originality, and integrity of the work.

\bibliographystyle{abbrvnat}
\bibliography{ref}

\clearpage
\appendix
\setcounter{table}{0}
\renewcommand{\thetable}{\textbf{A\arabic{table}}}
\renewcommand{\theHtable}{A\arabic{table}}
\setcounter{figure}{0}
\renewcommand{\thefigure}{\textbf{A\arabic{figure}}}
\renewcommand{\theHfigure}{A\arabic{figure}}

\section{Appendix Overview}
This appendix develops the execution details behind LoopDuet, then examines how task, context,
batching, and load affect depth allocation. Recurrent-state and FFN-routing studies explain the
choice of token roles and execution granularity. Section~\ref{app:tables} provides the complete
numerical comparisons.

\section{Workload and Implementation Details}
\label{app:implementation}
\paragraph{Notation and metrics.}
We write $(S,P)$ for decoding and prefilling core iterations, with $P\leq S$. We report task
accuracy, including exact match (EM) and token F1 for HotpotQA; accuracy differences are percentage
points. Joint-depth comparisons use $(32,32)$ as the full-depth baseline. Prefilling-depth
ablations compare $(S,P)$ with $(S,S)$.\footnote{Scores, differences, and ratios are
computed from unrounded values and rounded independently for display; subtracting or dividing
displayed values can therefore differ in the last digit.}

\paragraph{Workloads.}
GSM8K uses all 1,319 test examples, five demonstrations, greedy decoding, and extraction of the
final numeric answer. ARC-Challenge (ARC) uses all 1,172 test examples, five demonstrations, and the
highest-scoring choice label at the first answer position. HotpotQA uses all 7,405
distractor-development examples with normalized EM and token F1.

HotpotQA prompt construction is fixed within each model's comparisons. OLMo uses one demonstration
and four paragraphs: the two supporting passages and two distractors selected in their original
order. Distractors are removed as needed to satisfy the 1,000-token prompt cap. Huginn uses two
demonstrations and ten paragraphs. These inputs supply the supporting passages directly.

\paragraph{Models and execution.}
OLMo has 4/6/4 prelude/core/coda layers at width 2,048; Huginn has 2/4/2 at width 5,280.
Serving uses vLLM 0.14.1 on one A100-SXM4-80GB with bf16, an 85\% GPU-memory budget, and
prefix caching disabled. Uniform and split depths both use shared KV caches, zero recurrent-state
initialization, eager prefilling batches, and full CUDA graphs for decoding. Zero initialization
gives each serving token a fixed initial recurrent state.

A separate accuracy harness supports token-role and cache ablations. It uses the checkpoint's
random-state initialization and either shared or depth-indexed KV caches. In a depth-indexed
cache, skipped entries inherit the last computed recurrence at the same physical core layer.
Comparisons use the same implementation and cache layout for both configurations; serving accuracy
and performance come from the same vLLM runs.

\paragraph{Parallel tokens, sequential chunks and loops.}
Each core application processes the token positions in a scheduled prompt chunk together with
causal attention. Successive loops update their recurrent states, and successive chunks extend the
cached context. The final prompt position produces the first output token. Decoding then advances
one generated position per request at a time.

Algorithm~\ref{alg:depth} implements LoopDuet's compact execution. The first $P$ loops process all
$N$ scheduled tokens. The remaining $S-P$ loops operate on a gathered $B\times d$ tensor containing
each request's final scheduled query, where $B$ is the request count and $d$ the hidden width.
Every prompt-chunk endpoint continues to depth $S$, including intermediate endpoints, as do
decoding positions. The final prompt endpoint produces the first output token.

Prelude states, positions, and attention metadata follow the gathered indices. Scattering restores
the updated rows before the coda; other prompt positions retain their loop-$P$ context in one
shared KV plane per core layer. Batches whose maximum query length is one execute all $S$ loops.
\begin{algorithm}[t]
\caption{LoopDuet execution in one engine step}
\label{alg:depth}
\begin{algorithmic}[1]
\REQUIRE Scheduled tokens $X$, positions, KV cache; $S$, $P$
\STATE $E\leftarrow\operatorname{Prelude}(X)$; initialize recurrent state $H$
\IF{batch has multiple-query prefilling work and $P<S$}
  \FOR{$k=1$ to $P$}
    \STATE $H\leftarrow\operatorname{Core}(H,E,\mathrm{KV})$
  \ENDFOR
  \STATE $I\leftarrow$ last query position of each scheduled request
  \STATE $\widehat H\leftarrow\operatorname{Gather}(H,I)$; $\widehat E\leftarrow E_I$
  \STATE $\widehat M\leftarrow$ positions and attention metadata for $I$
  \FOR{$k=P+1$ to $S$}
    \STATE $\widehat H\leftarrow\operatorname{Core}(\widehat H,\widehat E,\mathrm{KV};\widehat M)$
  \ENDFOR
  \STATE $H\leftarrow\operatorname{Scatter}(H,I,\widehat H)$
\ELSE
  \FOR{$k=1$ to $S$}
    \STATE $H\leftarrow\operatorname{Core}(H,E,\mathrm{KV})$
  \ENDFOR
\ENDIF
\STATE \textbf{return} $\operatorname{Coda}(H)$
\end{algorithmic}
\end{algorithm}

\paragraph{Arrival and control semantics.}
Open-loop requests are timestamped by scheduled arrival and admitted before engine steps, so
TTFT includes admission delay and queueing. Alternating-rate traces use seeded exponential
interarrivals with a new rate every 30 seconds. Closed-loop runs replenish completed requests to
maintain at most $C$ active requests. In open-loop runs, $C=128$ limits the number of
sequences scheduled per engine step. Requests enter the engine when due and can wait in its
queue beyond this scheduling cap.

LoopDuet's queue controller observes the oldest first-token waiting age, including prefilling requests
already in service. It maps ages from 0.25 to 1 second linearly onto the configured depth range,
then rounds to an integer. The main configuration fixes $S=4$ and selects $P\in\{2,3,4\}$ before
each engine step. A request spanning several chunks can encounter different budgets.
The supplementary rate controller estimates demand from arrivals over a trailing window of at most
five seconds, plus the number waiting divided by a two-second drain interval. It selects the
deepest configured depth with sufficient measured capacity.

Controller occupancy measures the fraction of wall time spent at each prefilling depth
(Table~\ref{tab:control_occupancy}). A request's executed depth follows the policy in effect
when each prompt chunk is scheduled.

\paragraph{Timing measurements.}
We measure end-to-end serving time, excluding model loading, an eight-request warm-up, and
diagnostic calibration. Latency percentiles summarize requests within each run. The joint-depth
comparison uses one execution per configuration at $C=64$ and an 8,192-token engine-step budget.
Load and batching studies vary the arrival trace, concurrency cap, or token budget.

\paragraph{Depth updates and execution modes.}
Prefilling depth can be set at startup or changed in a running engine. At fixed $(S,P)=(11,4)$
and 40 requests/s, a runtime update in a reused engine yields EM 38.28 and median TTFT
0.15 seconds; startup configuration in a separate process yields EM 38.20 and 0.17 seconds.
Both use $C=128$ with the same prompts, arrival seed, cache, and precision.

Eager prefilling with full decoding CUDA graphs is faster than compiled execution for both
uniform-depth configurations: 304.7 versus 432.6 seconds for OLMo $(11,11)$, and 5,639.3 versus
6,877.2 seconds for Huginn $(24,24)$. Uniform-depth and LoopDuet comparisons use this faster
execution path.

\section{Joint Depth Selection and Context Sensitivity}
\label{app:selection}
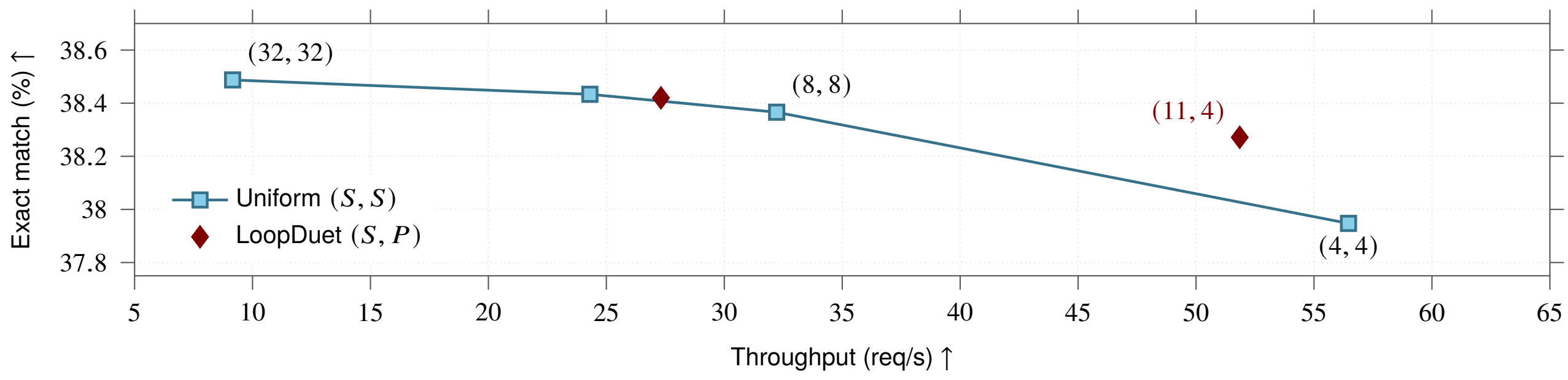
\begin{figure}[t]\centering
\begin{tikzpicture}\begin{axis}[paperaxis,width=.99\columnwidth,height=4.3cm,xmin=5,xmax=65,ymin=37.75,ymax=38.7,xlabel={Throughput (req/s) $\uparrow$},ylabel={Exact match (\%) $\uparrow$},label style={font=\plotfont},tick label style={font=\plotfont},scaled y ticks=false,ytick={37.8,38.0,38.2,38.4,38.6},legend style={font=\plotfont,draw=none,at={(.02,.05)},anchor=south west},axis lines=box]
\addplot[uniformline] table[x=req_s,y=em,col sep=comma]{figures/frontier_olmo32_uniform.csv};\addlegendentry{Uniform $(S,S)$}
\addplot[splitmarks] table[x=req_s,y=em,col sep=comma]{figures/frontier_olmo32_split.csv};\addlegendentry{LoopDuet $(S,P)$}
\node[font=\plotfont,anchor=south west] at (axis cs:9.156,38.49) {$(32,32)$};
\node[font=\plotfont,anchor=south west] at (axis cs:32.22,38.37) {$(8,8)$};
\node[font=\plotfont,anchor=south east,text=methodMaroon] at (axis cs:51.85,38.27) {$(11,4)$};
\node[font=\plotfont,anchor=north] at (axis cs:56.46,37.95) {$(4,4)$};
\end{axis}\end{tikzpicture}
\caption{\textbf{Uniform depth four supplies OLMo's starting budget.} Uniform $(4,4)$ has the highest throughput among the six HotpotQA settings, all of which satisfy the quality criterion relative to $(32,32)$. Table~\ref{tab:joint_all} reports F1 and EM changes.}\label{fig:olmo}\end{figure}
Joint depth selection exposes different allocations for the two models. Across 11 Huginn
configurations, LoopDuet's split budgets improve the accuracy--throughput choices highlighted in the main
paper. Across six OLMo configurations, uniform $(4,4)$ provides the highest throughput:
56.46 requests/s at EM 37.95, 0.54 points below $(32,32)$
(Figure~\ref{fig:olmo}). This setting supplies the starting budget for OLMo's controller. Selection uses the
closed-loop quality--throughput comparison; decoding depth is then fixed across the open-loop
arrival traces.

The useful prefilling depth also varies by task. In the accuracy harness, OLMo ARC $(11,4)$ and
$(32,4)$ change accuracy by $-1.54$ and $-1.88$ points relative to their same-$S$ controls.
Increasing $P$ to eight changes these deltas to $+0.85$ and $0.00$. On Huginn HotpotQA at $S=32$,
$P=8$ changes EM by $-2.63$ points, whereas $P=16$ changes it by $-0.23$.
Huginn ARC remains more sensitive: $(32,16)$ and $(14,8)$ change accuracy by $-1.45$ and $-4.44$
points. Depth allocation therefore depends on the task as well as the checkpoint.

Context length changes the accuracy cost even when runtime gains remain similar. Relative to
$(24,24)$, Huginn $(24,12)$ achieves approximately $1.85\times$ speedup with two, four, and ten
paragraphs, while EM changes by $-0.43$, $-0.45$, and $-1.92$ points. OLMo $(11,4)$ with two
paragraphs achieves $2.11\times$ speedup and a $-1.35$-point EM change relative to $(11,11)$.
Depth selection therefore uses task accuracy for each prompt construction. The state-distance
thresholds in Section~\ref{app:diagnostics} are calibrated on the default prompts.

\section{Serving Gains across Batch Budgets}
\label{app:batching}
\paragraph{Layer-pass savings translate into shorter serving times.}
On HotpotQA, reducing OLMo's prefilling depth from 11 to four at fixed $S=11$ lowers wall time
from 304.7 to 142.8 seconds and median TPOT from 0.420 to 0.196 seconds
(Table~\ref{tab:serving}).

The long-prompt limits $c_S/c_P$ are 3.57, 2.31, and 1.94 for OLMo $(32,8)$, OLMo $(11,4)$,
and Huginn $(32,16)$, respectively. A finite-workload estimate also accounts for the measured
mean prompt length $\bar L$ and generated-output count $\bar O$. Let $a,b,c$ count prelude,
core, and coda layers, so $c_k=a+bk+c$. The unchunked layer-pass estimate is
\[
 \widehat W(S,P)=(\bar L-1)c_P+\bar O c_S.
\]
The $\bar O$ full-depth evaluations comprise the final prompt position and the
$\bar O-1$ subsequent decoding positions. Dividing the uniform estimate by the LoopDuet
estimate gives 3.52, 2.29, and 1.94; measured speedups are 85--96\% of these values.
Measured runtime also includes work at additional chunk endpoints and non-layer costs.
The corresponding GSM8K estimates and timings appear in Table~\ref{tab:gsm8k_serving}.

\paragraph{Batch budgets change grouping while preserving the savings.}
Huginn's prefilling savings persist across batch sizes. At $C=128$, $(32,16)$ reaches
$1.84\times$, $1.87\times$, and $1.88\times$ speedup over $(32,32)$ with engine-step token budgets
of 4,096, 8,192, and 16,384. At the default 8,192-token budget, speedup remains $1.87\times$ for
$C=32,64,128$. For these long prompts, the token budget limits admitted prefilling work before the
request-concurrency cap. Table~\ref{tab:batch_budget} gives the engine-step counts and
latencies for the three token budgets.

The open-loop Huginn runs at four and eight requests/s are overload stress tests:
arrivals exceed the measured service capacities, so TTFT includes accumulated queueing.
At Poisson four requests/s, the 4,096- and 16,384-token comparisons reach $1.84\times$ and
$1.88\times$ speedup, with EM changes of $+0.41$ and $-0.35$ points.
Huginn $(24,12)$ versus $(24,24)$ reaches approximately $1.85\times$ speedup at both four and eight
requests/s, with EM changes of $-1.82$ and $-1.78$ points. Thus the runtime benefit persists under
queueing, while the selected depth determines the associated accuracy cost.

\section{Fixed-Depth Capacity and Queueing}
\label{app:queueing}
Reducing prefilling work can prevent backlog from accumulating at an otherwise unsustainable
arrival rate. For OLMo at $S=11$ and Poisson 40 requests/s, reducing $P$ from 11 to four lowers
median TTFT from 57.7 to 0.17 seconds and raises one-second attainment from 0.7\% to 100\%.
EM changes from 38.34 to 38.20. Crossing the required service capacity produces a larger latency
improvement than the execution-time ratio alone.

At 80 requests/s, both configurations accumulate backlog, but $P=4$ lowers median TTFT from
104.8 to 24.5 seconds. LoopDuet therefore reduces service work and expands the range of arrival
rates supported within a latency target.

\section{Quality Retention under Changing Load}
\label{app:controller}
\paragraph{Adaptation preserves depth when capacity permits.}
Starting from OLMo $(4,4)$, the controller fixes $S=4$ and adjusts $P\in\{2,3,4\}$.
At Poisson 60 requests/s and on the 10/60 burst trace, it attains one-second TTFT for every request
at EM 37.27 and 37.77. These scores exceed fixed $P=3$, the most accurate tested fixed depth with
the same attainment, by 0.92 and 1.35 points.

At 40 requests/s, available capacity permits $P=4$ throughout, so the queue controller meets
the latency target while retaining the starting depth. Fixed $P=4$ also attains 100\%;
the two policies yield EM 37.92 and 37.96, respectively.\footnote{The policies are executed
independently. Fixed $P=4$ uses 7,652 engine steps with median TTFT 0.087 seconds; the queue
policy uses 5,938 steps with median TTFT 0.112 seconds.}

Higher load makes the quality--latency trade-off explicit. On 20/80 bursts, adaptation gives
EM 35.12 at 87.5\% attainment, compared with EM 31.91 at 100\% for fixed $P=2$.
Retaining deeper computation gains 3.21 EM points while reducing target attainment by
12.5 percentage points.

\paragraph{The depth range determines the available response.}
At $S=11$, the queue policies use $(P_{\min},P_{\max})=(4,11)$, $(2,11)$, or $(2,4)$;
the rate policies use $(2,11)$ or $(4,11)$. Each variant retains its parameters across
its evaluated traces.

At 40 requests/s, allowing the queue controller to return to $P_{\max}=11$ gives
63.3\% attainment with $P_{\min}=4$ and 92.3\% with $P_{\min}=2$.
Setting $(P_{\min},P_{\max})=(2,4)$ gives 100\% at EM 38.35; fixed $(11,4)$ also
gives 100\%, at EM 38.20. This load can support deeper decoding than the main
$S=4$ configuration. The upper prefilling budget therefore affects queueing as well
as the amount of computation retained.

The rate controller illustrates the role of the lower budget (Table~\ref{tab:rate_control}).
Both $P_{\min}$ settings meet the target at 20 and 40 requests/s; static full depth at
20 requests/s gives 99.96\% attainment and EM 38.41. At 60 requests/s, $P_{\min}=2$ provides
enough capacity for 100\% attainment at EM 33.25, while $P_{\min}=4$ retains EM 38.11 with
4.1\% attainment.

Burst arrivals expose the same trade-off between service capacity and quality. On 20/80 bursts,
the queue controller with $P\in[2,4]$ retains EM 34.27 at 34.1\% attainment, while fixed
$P=2$ reaches 72.2\% attainment at EM 32.25.

\begin{table}[t]
\centering\small
\caption{Lower $P_{\min}$ increases capacity under heavy load. Rate-based control at
$S=11$, $P_{\max}=11$; rates are requests/s, and TTFT columns report one-second attainment.}
\label{tab:rate_control}
\begin{tabular}{@{}lrrrr@{}}
\toprule
& \multicolumn{2}{c}{$P_{\min}=2$} & \multicolumn{2}{c}{$P_{\min}=4$}\\
Arrival trace & TTFT (\%) & EM & TTFT (\%) & EM\\
\midrule
Poisson 20 & 100 & 38.74 & 100 & 38.70\\
Poisson 40 & 100 & 38.30 & 100 & 38.27\\
Poisson 60 & 100 & 33.25 & 4.1 & 38.11\\
Burst 10/60 & 99.9 & 34.23 & 23.7 & 38.10\\
Burst 20/80 & 44.6 & 34.33 & 15.0 & 38.28\\
\bottomrule
\end{tabular}

\end{table}

At 80 and 120 requests/s, offered load exceeds every tested policy's measured capacity at its
shallowest depth. The queue variants with $P_{\max}=11$ spend 97.6--98.8\% of wall time
at $P_{\min}$ (Table~\ref{tab:control_occupancy}), identifying loads that require
additional capacity beyond the available depth reduction.

\section{Recurrent-State Diagnostics and Task Sensitivity}
\label{app:diagnostics}
\paragraph{Replaying recorded answers exposes intermediate states.}
We first generate answers at each checkpoint's reference depth. We then run the model on the
original prompts and recorded answers, keeping the answer tokens fixed and recording hidden states
at every loop. Reference depths are 32 for OLMo r32 and Huginn, and 16 for OLMo r16. During replay,
next-token predictions match the recorded answer tokens at a rate of 95.1--99.6\% across the nine
checkpoint--task combinations.

For token $i$, the normalized distance between loops $P$ and $S$ is
$d_i(P;S)=\lVert h_i^P-h_i^S\rVert_2/\lVert h_i^S\rVert_2$.
One replay provides the intermediate states for every candidate loop pair. For example, OLMo
GSM8K gives 90th-percentile prefilling distances $q_{.9}(4;11)=0.180443$ and
$q_{.9}(4;32)=0.182533$. Huginn's states continue changing through loop 32, so its distances
describe deviation from this finite reference depth.

\paragraph{Role sensitivity extends beyond settling order.}
At median distance 0.1, Huginn HotpotQA crosses at loop 15 for ordinary prefilling tokens and
loop 19 for decoding tokens. On GSM8K, the crossings are 20 and 19. Earlier prefilling-state
settling is therefore task dependent.

Quality-based thresholds show a larger role difference for mean and upper-tail distances than for
the median. The fitted prefilling-to-decoding threshold ratios are 2.22 for the 90th percentile,
2.15 for the mean, and 1.11 for the median, using 44 prefilling-depth configurations and 28
uniform-depth configurations. These fits connect state changes to the quality impact of each
depth intervention.

\paragraph{Small state changes can affect tasks differently.}
On ARC, state distances below 0.18 accompany accuracy losses of 1.45--1.88 points
(Table~\ref{tab:state_sensitivity}). OLMo GSM8K $(32,8)$ likewise loses 1.29 points at distance
0.037, while $(32,4)$ has zero measured accuracy change at distance 0.183. Conversely, Huginn
GSM8K $(14,8)$ gains 0.68 points at distance 0.614. These observations suggest that the task
relevance of a state change matters alongside its magnitude, motivating task-specific depth
selection.

\begin{table}[t]
\centering\small
\caption{Task sensitivity varies at similar state-distance scales. The upper rows have small
distances with accuracy losses above one point; the lower rows retain accuracy within one point
at larger distances. Distances are prefilling-token 90th percentiles; changes use same-$S$ controls.}
\label{tab:state_sensitivity}
\begin{tabular}{@{}llrr@{}}
\toprule
Model / task & $(S,P)$ & Distance & $\Delta$ acc.\\
\midrule
Huginn / ARC & $(32,16)$ & 0.17201 & $-1.45$\\
OLMo r32 / ARC & $(11,4)$ & 0.11403 & $-1.54$\\
OLMo r32 / ARC & $(32,4)$ & 0.11518 & $-1.88$\\
OLMo r32 / GSM8K & $(32,8)$ & 0.03653 & $-1.29$\\
\midrule
Huginn / GSM8K & $(32,8)$ & 0.62001 & $-0.91$\\
Huginn / GSM8K & $(14,8)$ & 0.61445 & $+0.68$\\
Huginn / HotpotQA & $(14,8)$ & 0.46377 & $-0.72$\\
\bottomrule
\end{tabular}

\end{table}

\paragraph{Threshold transfer and aggregation.}
The diagnostic evaluation spans eight model--task groups across three checkpoints from two model
families. Depth pairs reuse states from the same replay. The pooled 90th-percentile threshold is
0.240964; fitting on two checkpoints and evaluating the third gives thresholds from 0.236799 to
0.240964 (Table~\ref{tab:diagnostic_transfer}).
We propose depths below the fitted threshold and evaluate their task accuracy.

For the binary comparisons in Tables~\ref{tab:diagnostic_transfer}--\ref{tab:diagnostic_selection}
and the complete diagnostic table, the quality criterion combines an accuracy loss of at most one
percentage point with a non-significant two-sided exact McNemar comparison at the five-percent
level. This paired test compares which examples change correctness against the same-$S$ control.
Agreement is the fraction of configurations for which the distance rule and this task-based
criterion give the same decision.

\begin{table}[t]
\centering\small
\caption{The fitted threshold remains near 0.24 when each checkpoint is held out in turn.
Each fit uses the other two checkpoints.}
\label{tab:diagnostic_transfer}
\begin{tabular}{@{}lrr@{}}
\toprule
Held-out checkpoint & Threshold & Agreement\\
\midrule
OLMo r32 & 0.240964 & 18/21\\
OLMo r16 & 0.240964 & 6/6\\
Huginn & 0.236799 & 13/17\\
\midrule
Combined & & 37/44\\
\bottomrule
\end{tabular}

\end{table}

The 90th-percentile rule retains its agreement on held-out checkpoints, whereas mean and
median aggregation give lower agreement after transfer (Table~\ref{tab:diagnostic_aggregation}).
All three aggregations match the per-model depth lookup in the pooled comparison.
The shared state-distance criterion supplies candidate budgets for task-accuracy validation.

Calibration can also cover both prefilling-depth and uniform-depth reductions with a common
threshold. A rule derived from decoding-token distances provides a parameter-free alternative,
with lower agreement on prefilling-depth choices. Table~\ref{tab:diagnostic_aggregation}
compares these calibration choices.

\begin{table}[t]
\centering\small
\caption{Upper-tail aggregation retains its agreement across checkpoints. Agreement is measured
against the task-accuracy criterion. The first five rules use 44 prefilling-depth configurations;
the joint rule adds 28 uniform-depth configurations. Held-out results combine the three checkpoint
folds; dashes mark settings without a held-out evaluation.}
\label{tab:diagnostic_aggregation}
\begin{tabular}{@{}lrr@{}}
\toprule
Selection rule & Pooled & Held-out\\
\midrule
90th-percentile distance & 37/44 & 37/44\\
Mean distance & 37/44 & 36/44\\
Median distance & 37/44 & 34/44\\
Per-model depth lookup & 37/44 & ---\\
Decoding-derived rule & 35/44 & ---\\
Joint 90th-percentile rule & 60/72 & ---\\
\bottomrule
\end{tabular}

\end{table}

\begin{table}[t]
\centering\small
\caption{State distance and task sensitivity jointly inform depth selection. Counts group
configurations by the pooled distance threshold and the measured quality criterion.
Table~\ref{tab:state_sensitivity} shows the task-level accuracy effects.}
\label{tab:diagnostic_selection}
\begin{tabular}{@{}lrr@{}}
\toprule
& \multicolumn{2}{c}{Quality criterion}\\
State distance & Satisfied & Not satisfied\\
\midrule
$q_{.9}\leq 0.240964$ & 18 & 4\\
$q_{.9}>0.240964$ & 3 & 19\\
\bottomrule
\end{tabular}

\end{table}

\paragraph{Prompt composition shapes the aggregate distance.}
Demonstrations account for 90.28\% and 91.10\% of ordinary prefilling tokens in OLMo r32 and
Huginn GSM8K, and approximately 82\% on ARC. Aggregate distances therefore largely reflect the
demonstration portion of these few-shot prompts.

\section{FFN Routing and Batch Unions}
\label{app:routing}
Batch savings from per-token sparsity depend on overlap among selected neurons. To characterize
this overlap and its quality cost, we select late-loop FFN neurons from earlier-loop activations
and apply masks before the down projection. The measurements cover task accuracy and neuron
coverage under dense execution.

At 50\% retention, accuracy changes by $-0.14$ points on OLMo HotpotQA, $-0.34$ on OLMo ARC, and
$+0.09$ on Huginn ARC. GSM8K changes by $-2.50$ for OLMo and $-2.73$ for Huginn; Huginn HotpotQA
changes by $-0.86$. A separate frozen-routing comparison uses OLMo r32 at $S=11$ on all
1,319 GSM8K examples: the first four loops are dense, and later loops retain a fixed
per-token subset selected from loop-four activations. Retaining 25\% gives 34.42\%
accuracy versus 47.61\% for its dense control, a 13.19-point decrease
(Table~\ref{tab:ffn_retention}). Static and grouped routes yield losses of 1.8--20.6 points. The quality cost thus depends on the
task, retained fraction, and routing granularity.

At batch size 16, the union of 50\% per-token selections covers 99.80--99.95\% of neurons
(Table~\ref{tab:neuron_union}).
Nearly every neuron remains active for some token in the batch. LoopDuet instead removes later
core applications from ordinary prefilling positions at a common boundary, making the reduction
explicit in the dense batched execution.

\section{Complete Comparison Tables}
\label{app:tables}
The tables
fixed-$S$ HotpotQA and GSM8K serving comparisons
(Tables~\ref{tab:serving} and~\ref{tab:gsm8k_serving}), batch-budget and controller-occupancy
measurements (Tables~\ref{tab:batch_budget} and~\ref{tab:control_occupancy}), and frozen
FFN-retention comparisons (Table~\ref{tab:ffn_retention}). The complete serving,
accuracy-harness, and state-distance tables follow
(Tables~\ref{tab:all_serving}--\ref{tab:all_diagnostics}), with neuron-union coverage in
Table~\ref{tab:neuron_union}.

Run identifiers encode task (\texttt{hp}, \texttt{arc}, or \texttt{gsm}; no task prefix in
accuracy runs denotes GSM8K), depths (\texttt{s}, \texttt{p}), shared KV (\texttt{kvsh}),
a shallow final prompt token (\texttt{ls}), and FFN routing (\texttt{sf}).
\texttt{C} is the sequence cap, \texttt{T} is the engine-step token budget, and
\texttt{pois} and \texttt{burst} identify arrival traces. Open-loop runs use $C=128$
even when the suffix is omitted; the default token budget is 8,192.
\texttt{hp2} and \texttt{hp4} indicate two- and four-paragraph context variants.
For $S=11$, \texttt{ctrl2}, \texttt{ctrl4}, and \texttt{ctrl42} denote queue ranges
$(P_{\min},P_{\max})=(2,11),(4,11),(2,4)$; \texttt{rate2} and \texttt{rate4}
use the corresponding lower bounds with $P_{\max}=11$.

\onecolumn
\small
\begin{longtable}{@{}llrrrrl@{}}
\caption{Complete measured joint-depth configurations on HotpotQA at $C=64$. Holds uses the full $(32,32)$ serving control.}\label{tab:joint_all}\\
\toprule Model & $(S,P)$ & Req/s & EM & F1 & $\Delta$EM & \\\midrule\endhead
Huginn & $(32,32)$ & 0.976 & 33.79 & 44.42 & +0.00 & \\
Huginn & $(24,24)$ & 1.313 & 33.52 & 44.19 & -0.27 & \\
Huginn & $(20,20)$ & 1.536 & 33.52 & 44.14 & -0.27 & \\
Huginn & $(18,18)$ & 1.698 & 32.72 & 43.19 & -1.07 & \\
Huginn & $(32,16)$ & 1.822 & 33.69 & 44.29 & -0.09 & \\
Huginn & $(16,16)$ & 1.897 & 32.07 & 42.40 & -1.72 & \\
Huginn & $(14,14)$ & 2.150 & 31.76 & 42.03 & -2.03 & \\
Huginn & $(24,12)$ & 2.431 & 31.60 & 41.91 & -2.19 & \\
Huginn & $(12,12)$ & 2.484 & 32.05 & 42.38 & -1.74 & \\
Huginn & $(14,8)$ & 3.489 & 30.33 & 40.35 & -3.46 & \\
Huginn & $(8,8)$ & 3.587 & 24.29 & 33.22 & -9.49 & \\
OLMo r32 & $(32,32)$ & 9.156 & 38.49 & 50.62 & +0.00 & \\
OLMo r32 & $(11,11)$ & 24.302 & 38.43 & 50.53 & -0.05 & \\
OLMo r32 & $(32,8)$ & 27.315 & 38.42 & 50.47 & -0.07 & \\
OLMo r32 & $(8,8)$ & 32.220 & 38.37 & 50.42 & -0.12 & \\
OLMo r32 & $(11,4)$ & 51.850 & 38.27 & 50.12 & -0.22 & \\
OLMo r32 & $(4,4)$ & 56.464 & 37.95 & 49.63 & -0.54 & \\
\bottomrule\end{longtable}

\begin{table}[t]
\centering\small
\setlength{\tabcolsep}{4pt}
\caption{Full HotpotQA development set, $C=64$, one A100-80GB. Each uniform-depth baseline uses $(S,S)$ with the same model, prompt format, cache, and engine path. Fraction is measured speedup divided by the analytic layer-pass ratio, rather than hardware utilization.}
\begin{tabular}{@{}llrrrrr@{}}
\toprule
Model / HotpotQA context & $(S,P)$ & \shortstack{Wall time (s)\\Uniform $\to$ LoopDuet} & Speedup & \shortstack{EM\\Uniform $\to$ LoopDuet} & Layer ratio & Fraction\\
\midrule
OLMo r32 / 4 paragraphs & $(32,8)$ & $808.7\to271.1$ & $2.98\times$ & $38.49\to38.42$ & 3.51 & 85\%\\
OLMo r32 / 4 paragraphs & $(11,4)$ & $304.7\to142.8$ & $2.13\times$ & $38.43\to38.27$ & 2.29 & 93\%\\
Huginn / 10 paragraphs & $(32,16)$ & $7588\to4064$ & $1.87\times$ & $33.79\to33.69$ & 1.94 & 96\%\\
\bottomrule
\end{tabular}
\label{tab:serving}
\end{table}

\begin{longtable}{@{}llrrrr@{}}
\caption{GSM8K serving comparisons on all 1,319 test examples at $C=64$. Wall-time pairs are uniform $(S,S)$ followed by LoopDuet $(S,P)$; accuracy changes are percentage points. Layer ratios use the same workload-dependent estimate as Table~\ref{tab:serving}.}\label{tab:gsm8k_serving}\\
\toprule Model & $(S,P)$ & Wall time (s) & Speedup & Layer ratio & $\Delta$ accuracy\\
\midrule\endhead
Huginn & $(14,8)$ & $151.43\to152.42$ & $0.99\times$ & 1.52 & $+0.68$\\
Huginn & $(32,16)$ & $324.94\to331.10$ & $0.98\times$ & 1.71 & $+2.12$\\
OLMo r32 & $(11,4)$ & $87.09\to86.34$ & $1.01\times$ & 2.00 & $+0.53$\\
OLMo r32 & $(32,8)$ & $225.61\to223.35$ & $1.01\times$ & 2.71 & $-0.15$\\
\bottomrule\end{longtable}

\begin{longtable}{@{}rrrrr@{}}
\caption{Huginn HotpotQA batching sweep at $C=128$: $(32,16)$ versus $(32,32)$. Steps count LoopDuet engine calls; median TTFT pairs are uniform followed by LoopDuet. Accuracy changes are EM percentage points.}\label{tab:batch_budget}\\
\toprule Token budget & Engine steps & Speedup & $\Delta$EM & Median TTFT (s)\\
\midrule\endhead
4,096 & 3,797 & $1.84\times$ & $+0.12$ & $126.6\to68.7$\\
8,192 & 1,902 & $1.87\times$ & $-0.09$ & $116.0\to62.1$\\
16,384 & 957 & $1.88\times$ & $-0.14$ & $103.1\to54.8$\\
\bottomrule\end{longtable}

\begin{longtable}{@{}rllrr@{}}
\caption{Queue-controller depth allocation over time. Each share is the percentage of total wall time spent at the indicated prefilling depth; time before the first controller update is unassigned. Requests follow the policy active when their chunks are scheduled.}\label{tab:control_occupancy}\\
\toprule $S$ & $(P_{\min},P_{\max})$ & Arrival trace (req/s) & Depth $P$ & Wall time (\%)\\
\midrule\endhead
4 & $(2,4)$ & Burst 10/60 & 4 & 86.99\\
4 & $(2,4)$ & Burst 10/60 & 3 & 12.92\\
11 & $(2,11)$ & Poisson 80 & 2 & 97.62\\
11 & $(2,11)$ & Poisson 120 & 2 & 98.15\\
11 & $(4,11)$ & Poisson 80 & 4 & 98.73\\
11 & $(4,11)$ & Poisson 120 & 4 & 98.82\\
\bottomrule\end{longtable}

\begin{longtable}{@{}lrrr@{}}
\caption{Frozen FFN routing on OLMo r32, $S=11$, all 1,319 GSM8K examples. Loops 1--4 are dense; loops 5--11 retain the indicated fraction of neurons using a per-token selection frozen from loop four. The dense control uses all neurons throughout.}\label{tab:ffn_retention}\\
\toprule Retained neurons & Correct answers & Accuracy (\%) & $\Delta$ accuracy (pp)\\
\midrule\endhead
100\% (dense) & 628 & 47.61 & $+0.00$\\
50\% & 622 & 47.16 & $-0.45$\\
25\% & 454 & 34.42 & $-13.19$\\
\bottomrule\end{longtable}

\begin{longtable}{@{}lrrrrrr@{}}
\caption{Serving performance across depth, load, context, batching, and execution-mode settings (133 runs). TTFT is in seconds. Accuracy is EM on HotpotQA and numeric-answer correctness on GSM8K.}\label{tab:all_serving}\\
\toprule
Run & $n$ & Wall (s) & Acc. & TTFT p50 & TTFT p99 & $\leq1$s (\%) \\
\midrule\endfirsthead
\caption[]{Serving performance across depth, load, context, batching, and execution-mode settings (133 runs). TTFT is in seconds. Accuracy is EM on HotpotQA and numeric-answer correctness on GSM8K.}\\
\toprule
Run & $n$ & Wall (s) & Acc. & TTFT p50 & TTFT p99 & $\leq1$s (\%) \\
\midrule\endhead
\parbox[t]{.40\textwidth}{\raggedright\scriptsize olmo32\_s11\_ctrl2\_burst\_C128} & 7405 & 225.7 & 36.06 & 0.979 & 1.191 & 57.2 \\
\parbox[t]{.40\textwidth}{\raggedright\scriptsize olmo32\_s11\_ctrl2\_pois120\_C128} & 7405 & 98.9 & 32.10 & 18.554 & 36.247 & 0.3 \\
\parbox[t]{.40\textwidth}{\raggedright\scriptsize olmo32\_s11\_ctrl2\_pois40\_C128} & 7405 & 187.3 & 37.99 & 0.829 & 1.087 & 92.3 \\
\parbox[t]{.40\textwidth}{\raggedright\scriptsize olmo32\_s11\_ctrl2\_pois80\_C128} & 7405 & 99.3 & 32.21 & 3.120 & 5.768 & 0.7 \\
\parbox[t]{.40\textwidth}{\raggedright\scriptsize olmo32\_s11\_ctrl42\_burst} & 7405 & 225.3 & 37.25 & 0.580 & 0.939 & 100.0 \\
\parbox[t]{.40\textwidth}{\raggedright\scriptsize olmo32\_s11\_ctrl42\_burst2080} & 7405 & 162.5 & 34.27 & 1.278 & 2.455 & 34.1 \\
\parbox[t]{.40\textwidth}{\raggedright\scriptsize olmo32\_s11\_ctrl42\_pois40} & 7405 & 186.5 & 38.35 & 0.150 & 0.354 & 100.0 \\
\parbox[t]{.40\textwidth}{\raggedright\scriptsize olmo32\_s11\_ctrl42\_pois60} & 7405 & 125.1 & 36.93 & 0.648 & 0.975 & 99.6 \\
\bottomrule
\end{longtable}

\begin{longtable}{@{}llrrrr@{}}
\caption{All 169 supplied paired accuracy comparisons; accuracy and deltas in percentage points.}\\
\toprule
Run & Control & $n$ & Control acc. & Run acc. & $\Delta$ \\
\midrule\endhead
\parbox[t]{.25\textwidth}{\raggedright\scriptsize arc\_huginn\_s14\_p14} & \parbox[t]{.25\textwidth}{\raggedright\scriptsize arc\_huginn\_s24\_p24} & 1172 & 43.52 & 38.82 & -4.69 \\
\parbox[t]{.25\textwidth}{\raggedright\scriptsize arc\_huginn\_s14\_p14} & \parbox[t]{.25\textwidth}{\raggedright\scriptsize arc\_huginn\_s32\_p32} & 1172 & 43.60 & 38.82 & -4.78 \\
\parbox[t]{.25\textwidth}{\raggedright\scriptsize arc\_huginn\_s14\_p14\_sf0.5s8} & \parbox[t]{.25\textwidth}{\raggedright\scriptsize arc\_huginn\_s14\_p14} & 1172 & 38.82 & 38.91 & +0.09 \\
\parbox[t]{.25\textwidth}{\raggedright\scriptsize arc\_huginn\_s14\_p4} & \parbox[t]{.25\textwidth}{\raggedright\scriptsize arc\_huginn\_s14\_p14} & 1172 & 38.82 & 24.66 & -14.16 \\
\parbox[t]{.25\textwidth}{\raggedright\scriptsize arc\_huginn\_s14\_p4} & \parbox[t]{.25\textwidth}{\raggedright\scriptsize arc\_huginn\_s4\_p4} & 1172 & 23.46 & 24.66 & +1.19 \\
\parbox[t]{.25\textwidth}{\raggedright\scriptsize arc\_huginn\_s14\_p8} & \parbox[t]{.25\textwidth}{\raggedright\scriptsize arc\_huginn\_s14\_p14} & 1172 & 38.82 & 34.39 & -4.44 \\
\parbox[t]{.25\textwidth}{\raggedright\scriptsize arc\_huginn\_s14\_p8} & \parbox[t]{.25\textwidth}{\raggedright\scriptsize arc\_huginn\_s8\_p8} & 1172 & 28.75 & 34.39 & +5.63 \\
\parbox[t]{.25\textwidth}{\raggedright\scriptsize arc\_huginn\_s14\_p8\_sf0.5s8} & \parbox[t]{.25\textwidth}{\raggedright\scriptsize arc\_huginn\_s14\_p14} & 1172 & 38.82 & 34.47 & -4.35 \\
\bottomrule
\end{longtable}

\begin{longtable}{@{}llrrrrll@{}}
\caption{Prefilling-state distance and task accuracy across 44 depth configurations. Distances compare loops $P$ and $S$ from the same reference-depth replay. The quality criterion summarizes measured accuracy; the distance threshold identifies candidate depths.}\label{tab:all_diagnostics}\\
\toprule
Model & Task & $S$ & $P$ & Distance & $\Delta$ &  &  \\
\midrule\endfirsthead
\caption[]{Prefilling-state distance and task accuracy across 44 depth configurations. Distances compare loops $P$ and $S$ from the same reference-depth replay. The quality criterion summarizes measured accuracy; the distance threshold identifies candidate depths.}\\
\toprule
Model & Task & $S$ & $P$ & Distance & $\Delta$ &  &  \\
\midrule\endhead
Huginn & ARC & 14 & 4 & 0.93704 & -14.16 &   &   \\
Huginn & ARC & 14 & 8 & 0.50296 & -4.44 &   &   \\
Huginn & ARC & 32 & 4 & 0.93917 & -19.37 &   &   \\
Huginn & ARC & 32 & 8 & 0.51466 & -8.36 &   &   \\
Huginn & ARC & 32 & 16 & 0.17201 & -1.45 &   &  \\
Huginn & GSM8K & 14 & 4 & 1.03466 & -11.14 &   &   \\
Huginn & GSM8K & 14 & 6 & 0.80493 & -2.20 &   &   \\
Huginn & GSM8K & 14 & 8 & 0.61445 & +0.68 &  &   \\
\bottomrule
\end{longtable}

\begin{longtable}{@{}llrrrrrr@{}}
\caption{At batch size 16, per-token FFN sparsity leaves nearly all neurons active.
Neuron-union coverage (percent) for 50\% per-token routing.}
\label{tab:neuron_union}\\
\toprule
Model & Task & $S$ & Answer tokens & $B=4$ & $B=16$ & $B=64$ & Groups of 32, $B=16$\\
\midrule\endhead
OLMo r32 & GSM8K & 11 & 116,390 & 91.35 & 99.95 & 100.00 & 99.72 \\
OLMo r32 & HotpotQA & 11 & 28,173 & 90.02 & 99.85 & 100.00 & 98.29 \\
Huginn & GSM8K & 14 & 148,075 & 90.50 & 99.87 & 100.00 & 97.61 \\
Huginn & HotpotQA & 32 & 27,952 & 89.52 & 99.80 & 100.00 & 97.16 \\
\bottomrule
\end{longtable}

\end{document}